\documentstyle[12pt,apjpt4]{article}
\input{psfig}
\renewcommand{\baselinestretch}{1}

\begin{document}
\begin{titlepage}

\title{ \huge New Insights from HST Studies of Globular Cluster Systems I:  Colors, Distances and Specific Frequencies of 28 Elliptical Galaxies \altaffilmark{1}    }
\vspace*{0.5 truein}
\author{ Arunav Kundu \altaffilmark{2} }
\vspace*{0.2 truein}
\affil{ Space Telescope Science Institute, 3700 San Martin Drive, Baltimore, MD 21218 }
\vspace*{0.2 truein}
\affil{  Electronic Mail: akundu@astro.yale.edu}
\vspace*{0.2 truein}
\affil{ and }
\vspace*{0.2 truein}
\affil{ Dept of Astronomy, University of Maryland, College Park, MD 20742-2421 }
\vspace*{0.5 truein}
\author{ Bradley C. Whitmore }
\vspace*{0.2 truein}
\affil{ Space Telescope Science Institute, 3700 San Martin Drive, Baltimore, MD 21218 }
\vspace*{0.2 truein}
\affil{  Electronic Mail: whitmore@stsci.edu }

\altaffiltext{1}{\normalsize Based on observations with the 
NASA/ESA {\it Hubble Space Telescope}, obtained from the data Archive at the Space Telescope Science 
Institute, which is operated by the Association of Universities for Research 
in Astronomy, Inc., under NASA contract NAS5-26555 }

\altaffiltext{1}{\normalsize Present address:  Astronomy Department, Yale University, 260 Whitney Av., New Haven, CT 06511} 

 \begin{abstract}

	We present an analysis of the globular cluster systems of 28 elliptical 
galaxies using archival WFPC2 images in the V and I-bands. The V-I color distributions of at least 50\% of the galaxies appear to be bimodal {\it at the present level of photometric accuracy}. We argue that this is indicative of multiple epochs of cluster formation early in the history of these galaxies, possibly due to mergers. We also present the
first evidence of bimodality in low luminosity galaxies and
discuss its implication on formation scenarios. The mean color of the 28 cluster systems studied by us is V-I = 1.04$\pm$0.04 (0.01) mag corresponding to a mean metallicity of Fe/H = -1.0$\pm$0.19 (0.04). We find that the turnover magnitudes of the globular cluster luminosity functions (GCLF) of our sample  are in excellent agreement with the distance measurements using other methods and conclude that the accuracy of the GCLF is at least as good as
the surface brightness fluctuation method. The absolute magnitude of the
 turnover luminosity of the GCLF is M$_V^0$ = -7.41 (0.03) in V  and M$_I^0$ = -8.46 (0.03) in I. The mean local specific frequency of our sample of elliptical galaxies within the WFPC2 field-of-view is 2.4$\pm$1.8 (0.4), considerably higher than the 1.0$\pm$0.6 (0.1) derived for a comparable sample of S0s in a similar analysis. It shows no obvious correlation with metallicity, host galaxy mass or membership in a galaxy cluster. The median size of clusters in all galaxies appears to be remarkably constant at $\sim$2.4 pc. We suggest that in the future it might be possible to use  the sizes of
clusters in the inner regions of galaxies as a simple geometrical distance indicator. 

\end{abstract}
\keywords { galaxies: distances and redshifts --- galaxies: elliptical and lenticular, cD --- galaxies: formation --- galaxies: general --- galaxies: star clusters }

\end{titlepage}

\section{Introduction}

	Globular clusters (GC) are among the most pristine objects in a galaxy, and are excellent fossil records of the chemical and dynamical properties of the
ambient gas in the era of their formation. As present observations suggest that most
clusters are formed during major episodes of star formation associated with a
significant event in the formation process of a galaxy, the study of globular
cluster systems (GCS) provides vital clues to the formation (and possibly evolutionary) history of the host.  
 
 The advent of the HST with its superior angular resolution has given a major boost to globular cluster research because of the ease with which cluster candidates can be identified.  While most HST-based papers concentrate on a single galaxy, or just a handful, there are very few integrated analyses that study a  large number of galaxies simultaneously (Forbes et al. 1996; Gebhardt \& Kissler-Patig  1999). Such analyses have the advantage of eliminating the possible systematic
 differences that make intercomparison of the results of different authors a
somewhat tricky task. The HST archives contain a number of high resolution images of galaxies
that provide a rich source of data for  a large scale survey. In this study, we analyse the GCSs of 28 elliptical galaxies observed by the WFPC2 in the V (F555W) and I (F814W) filters. A few of
these observations were made expressly for the purpose of studying globular
clusters, while most were observed for entirely different scientific reasons. 

Elliptical galaxies are ideally suited for GCS studies. With little or no dust obscuration and no complicated regions of star formation
it is relatively easy to identify globular clusters against the smooth
 background of the host. Moreover, ellipticals as a class have a much 
larger cluster density than S0's or spirals, yielding much better statistics. In a companion paper Kundu \& Whitmore (2001) (hereafter referred to as Paper II) we present a corresponding study of S0 galaxies.

\section{Observations and Data Reduction}

	 This analysis is based on  archival HST WFPC2 images of nearby elliptical galaxies. While the HST archive is teeming with deep images of individual galaxies in a single filter, the number of candidates
with color information (i.e. images in two or more filters) is substantially
 less. Although it is possible to learn much about globular clusters
from single images, color information  greatly enhances the 
scientific value. Therefore, we have chosen to study only candidate galaxies with at least two broad-band color images. For a variety of scientific
and instrumental reasons a large fraction of  WFPC2 observational programs image galaxies in  F555W (V) and 
F814W (I).  In this paper we have analyzed the cluster systems of 28 
galaxies observed in these two filters.

  The list of 28 program elliptical galaxies (and NGC 4550) along with
some salient information is presented  in Table 1.  Note that although NGC 4550 is a S0 galaxy it has been analysed here instead of Paper II as the reduction and analysis techniques applied to this set of deep images is similar to that of the ellipticals in this paper. In Paper II we study a homogeneous set of short exposure 'snapshot' S0 images, taking into consideration the special problems associated with such images. Wherever we compare the properties of ellipticals with S0s, e.g. the mean color or specific frequency, we take care to exclude the
NGC 4550 values from the elliptical sample, and to include it with the S0s.
 In Table 2 we list the date of the observations, the exposure times and the gains.

 We used the STSDAS utility GCOMBINE and the IRAF task COSMICRAYS  to remove
 cosmic rays and hot pixels respectively from the images in each filter.
 We then identified  cluster candidates 
using the technique described in Kundu \& Whitmore (1998). For this set of images we used a
 S/N cutoff of 3  to detect candidate clusters. We rejected sources that
did not satisfy the concentration criteria
2 $<$ $\frac{Counts_{3pix}}{Counts_{0.5pix}}$ $<$ 13 for the PC and 2 $<$
 $\frac{Counts_{3pix}}{Counts_{0.5pix}}$ $<$ 10 for the WF to weed out chip 
defects and background galaxies.   
Cluster candidates that satisfied the selection criteria in both the F555W and
the F814W image made it to our candidate source list for each galaxy.

  After correcting for the geometrical distortion (Holtzman et al.  1995a),
we performed aperture photometry  using a 3 pixel radius 
aperture for the PC and a 2 pixel radius  aperture for the WF, using the
median pixel value between 5 and 8 pixels as the sky background. 
 Since the profiles
of the cluster candidates are slightly broader than a stellar PSF, we derived
 the aperture correction by computing the flux within apertures of various sizes
 on a sample of bright clusters in $\sim$10 galaxies and compared these with
profiles of stars in the calibration field NGC 5139 ($\omega$ Cen) and the 
standard star GRW+70D5824. Due to the slightly resolved nature of the clusters, the magnitude of the aperture
 correction for the clusters candidates is always larger than the point source
correction reported by Holtzman et al.  (1995a) and varies inversely with the
 distance to a particular galaxy.
 We fit this distance dependent aperture correction to a straight line and derived the following relationships:

$\Delta$$_{ap(PC:F555W)}$ = -0.559 + 0.00929 * d  

$\Delta$$_{ap(PC:F814W)}$ = -0.588 + 0.00760 * d  

$\Delta$$_{ap(WF:F555W)}$ = -0.491 + 0.00598 * d  

$\Delta$$_{ap(WF:F814W)}$ = -0.508 + 0.00540 * d  

with limiting minimum (in absolute terms) possible values of:

 ($\Delta$$_{ap(PC:F555W)}$ $=$ -0.260)

 ($\Delta$$_{ap(PC:F814W)}$ $=$ -0.388)

 ($\Delta$$_{ap(WF:F555W)}$ $=$ -0.283)

 ($\Delta$$_{ap(WF:F814W)}$ $=$ -0.364)

where $\Delta$$_{ap}$ is the aperture correction in magnitudes, and d is the distance to the galaxy in Mpc. As the minimum (absolute) aperture correction for the unresolved cluster systems of distant galaxies cannot be smaller than 
that for a point source, whenever the first set of equations above return values smaller than those for unresolved sources (second set of equations above) we set the
aperture correction to the point source values. Note that the aperture corrections listed above  include the 0.1 mag correction from a standard 
aperture of 0.5$"$ to a nominal infinite aperture recommended by Holtzman et al.
 (1995b), consequently the zero points reported below do not include this factor.

For the PC we adopted photometric
zero points of 22.573 mag and 21.709 (following Whitmore et al.  1997) and small
 color correction terms from 
Holtzman et al.  (1995b) to convert the F555W and F814W magnitudes 
to Johnson V and  Cousins I respectively. These zero points - for a gain of 7 - are in
 excellent agreement with the values of Holtzman et al. (1995b), 22.56 in V and 21.69
in I, although they differ slightly from the more recent numbers reported in the {\it HST Data Handbook} (1997), 22.545 in F555W and 21.639 in F814W. In order to maintain
 consistency with our previous projects we chose to use the Whitmore et al. (1997) zero point.  Since the zero points for each of the WF 
chips are offset slightly from the PC values we added the small differences 
quoted in the {\it HST data handbook} (1997)  before applying the color correction 
terms to the candidate objects in the WF. For images with gains of 14 we further added
the corrective terms recommended in the Data Handbook.

   We corrected for the foreground
Galactic extinction in the direction of each of the observed galaxies using
 the A$_B$ values quoted in the NED (The NASA/IPAC Extragalactic Database)
 database and the reddening curve 
from Mathis (1990). The V-band extinctions are listed in Table 1 (Note that when we initiated this project the Schlegel at al. (1998) extinction curves were not available on the NED database, hence the values used in this paper are all based on the older Burstein \& Heiles (1982) compilation). An inspection of the V-I maps of the hosts revealed no significant regions of 
dust, hence we have ignored internal reddening. We did
 not correct for the Charge Transfer  Efficiency (CTE) gradient across the chips
  as the CTE problem is expected to be minimal in the presence of the 
strong galaxy background in most of our sample  objects. (Holtzman et al.  
1995b; Whitmore \& Heyer 1997). Even in regions
of low background the CTE corrections are expected to be very small ($\sim$0.01
 mag) at the epoch of these observations (Whitmore, Heyer, \& Casertano 1999).

	Ground-based identifications of point-like sources around
galaxies are usually contaminated by compact background sources and
foreground Galactic stars. Most ground-based analyses account for this by 
statistically subtracting the count of point-like sources in
$"$background$"$ images of a region of sky near the galaxy. While  we do not have the luxury of off-galaxy images,
 the superior angular resolution of the HST makes this almost redundant  since 
globular clusters can generally be distinguished both from unresolved foreground
 stars and background galaxies in WFPC2 images.

\section{Results and Discussion}

\subsection{Colors and Metallicity Distributions}

	The (V-I) vs V color-magnitude diagrams   of the point-like objects in the elliptical
sample are shown in Fig 1. The galaxies have been sorted by absolute total magnitude (Note: Throughout this paper we sort figures and tables by the absolute magnitude of the host, except in sections where we address the GCLF and sizes where we sort by the distance moduli from Table 1).   The  majority of  cluster candidates 
  lie in a narrow range of color  between 0.5$<$V-I$<$1.5 with 
a mean color near V-I $\approx$ 1.0 mag, which is fairly typical for old globular cluster systems.  In order to  filter 
out most of the few remaining contaminating foreground stars and background galaxies we
 shall consider only objects within the color range 0.5$<$V-I$<$1.5 to be bona fide
cluster candidates and assume that all objects outside this range are foreground and/or background contaminants. We visually scrutinized the candidates outside the selected color range and satisfied ourselves that  these (few) objects are likely to be foreground stars or compact galaxies. We also inspected the spatial distribution of the cluster candidates in each 
of the sample galaxies and found that in each case the density distribution was roughly centered on the nucleus of the 
galaxy, further confirming
 that they are bona fide  members of the GCS.

	The mean colors of the globular cluster candidates in the color range 
 0.5$<$V-I$<$1.5 and the corresponding metallicities, based on  the Galactic color-metallicity transformation equations reported in Kundu \& Whitmore (1998), are listed in Table 3.  The mean color of the GCSs of our entire elliptical sample is 1.04$\pm$0.04 (0.01) which is  indistinguishable from the value of
1.00$\pm$0.07 (0.01) for the S0s studied in Paper II. (Note - Throughout this
 paper we have used the following convention in quoting uncertainties: The number following the $\pm$ sign is the standard deviation, while a number in parentheses refers to the uncertainty in the mean.)
Therefore, inasmuch as the broad-band colors trace metallicities, we can conclude that the metallicity of the globular cluster systems of early type galaxies is not a function of Hubble type. The metallicities in Table 3 cover a fairly large range of values. What then drives this spread? In Paper II we find that the mean metallicities  of S0 galaxy GCSs are 
correlated to the galaxy luminosity (mass). It has been suggested that elliptical
galaxies follow a similar relationship. To check the veracity of this claim we plot the mean color of the elliptical GCSs vs the absolute magnitude of the host galaxies in Fig 2. For comparison we have also marked the S0 sample from Paper II. As for the S0s, we find that the metallicities of GCSs generally increase with luminosity, and that there is no offset between the the S0 and elliptical samples. While there are no low luminosity S0's or ellipticals with high mean metallicities (However see $\S$3.2), the
more luminous galaxies present an interesting phenomenon. There appear to be a number of luminous elliptical galaxies that have metallicities significantly 
lower than that expected for a linear [Fe/H]-M$_V^T$ relationship. Understanding this spread in GCS metallicities with
increasing luminosity may hold the key to understanding the bigger 
problem of globular cluster system (and galaxy) formation. 
This  difference in the mean metallicities of the cluster systems of hosts with similar luminosities may
  either be due to an offset in the entire metallicity distribution of the two 
cluster systems, or differing fractions of two or more sub-populations having 
different metallicities. To this end it is critical to study the detailed color 
distributions of individual GCSs. 

In Fig 3 we plot the histograms of the GC 
color distributions of the 28 elliptical galaxies (plus NGC 4550), sorted by the absolute integrated magnitude of the galaxy from Table 1. It is immediately obvious from  Figs 1 and 3 that some galaxies, e.g. NGC 4472, NGC 4649
 and NGC 4494,  have bimodal color distributions. Quite a few of the other
cluster systems also appear to be good candidates for bimodality. In order to objectively test for multiple modes we applied the KMM mixture modeling algorithm
 of Ashman, Bird \& Zepf (1994)  to the unbinned  distributions. We considered only those galaxies which showed a statistically convincing evidence
of bimodality in the KMM test (at better than the 95$\%$ confidence level) to have bimodal color distributions. KMM tests, revealed that 18 of our 29 sample galaxies (including M87 from Kundu et al. 1999) can be better described by two Gaussian sub-populations than a single Gaussian, while the other 11 showed no reasonable partitions for two or more groups. None of the GCSs show meaningful partitions for three or more groups. We also  visually inspected the histograms and color-magnitude diagrams of the GCSs and applied an even more stringent visual criteria to the bimodal candidate list to isolate 9 galaxies (again including M87) with very obvious bimodal color distributions. We suggest that these 9 galaxies certainly have bimodal GCS color distributions and the other
candidates for bimodality (from the KMM tests) are very likely to be bimodal.

 The likelihood of the color distributions of individual galaxies being bimodal and the positions of the Gaussian peaks of  the red and blue clusters calculated  by the KMM algorithm are listed in Table 3. Thus we can conclude that at least
30$\%$ and up to 60$\%$ of elliptical galaxies have bimodal color distributions.
Even this value of 60$\%$ may be an underestimate for a couple of reasons. In most galaxies with  bimodal color distributions studied to
date the difference in the peaks is of the order of V-I$\approx$0.2 mag. The evidence for bimodality may be suppressed in cases with low number statistics, significant photometric errors or smaller separations of the peak. Ashman et al.  (1994) show that for sample sizes of 100-200 objects the photometric 
uncertainties should be a factor of $\sim$2.5-3.0 less than the expected separation
between the peaks for the KMM algorithm to reliable detect or reject bimodality.
A more surreptitious problem may be that of the  age-metallicity conspiracy. One of the galaxies in our sample that shows no evidence for bimodality is NGC 3610. Whitmore et al.  (1997) showed that in this
merger remnant, the ages and metallicities of the clusters may be conspiring to produce a single peak with an extended blue tail. We also note that the bimodality fraction in our sample is somewhat higher than that reported by Gebhardt \& Kissler-Patig (1999), who found 7 out of the 16 galaxies with over a 100 clusters in their study to be bimodal. However we note that  typically our routines detect a 50\% higher number of
 clusters in galaxies in which the two studies use  the same HST data
 sets (Table 3). This is most likely due to the more
 stringent cluster selection/detection criteria imposed by them, which leads to fewer candidates in their sample. Consequently, we have better number statistics for bimodality tests. Based on the discussion in this section estimate (conservatively) that at least 50$\%$ of ellipticals show bimodal color distributions at the present level of photometric accuracy. 

\subsection {Implications on Formation Models}
	There are three competing models/scenarios that attempt
to explain the  bimodality in the cluster colors. While the merger model (Schweizer 1987; Ashman $\&$ Zepf 1992) and the  multiple collapse model (Forbes, Brodie, \& Grillmair 1997) both suggest that this is the consequence of globular clusters forming during two distinct epochs in the metal enrichment histories of these galaxies, the C\^{o}t\'{e} et al.  (1998) model proposes that all cluster systems formed at 
roughly the same time and that the bimodal distribution is a result of the merger (without creation of new clusters) or cannibalism of a metal-poor GCS by a larger galaxy with a pre-existing metal-rich GCS. 
Ashman $\&$ Zepf (1992) and Forbes et al. (1997) differ in the mechanism of  formation of the younger metal-rich clusters. While the merger model suggests creation of metal-rich GCs during a major interaction,
the multiple collapse model proposes that the second burst is triggered during
a secondary collapse phase of the galaxy (although  Forbes et al. do not suggest a trigger mechanism).

	One of the most interesting results from Fig 3 and Table 2 is that there is evidence for bimodality in the GCSs of galaxies at all luminosities.
This is a fairly  important discovery which has  significant implications on
the formation models. In the past, the lack of bimodality in low luminosity ellipticals and their alleged narrow color distributions has been used to argue 
that they might not have formed in a major merger of stellar systems, or to restrict the merger parameters on the basis of a color-metallicity conspiracy (Kissler-Patig, Forbes \& Minniti
1998). A case in point is the cluster system of NGC 1427. Kissler-Patig et al. 
 (1998) claim, partly on the basis of the WFPC2 observations of Forbes et al.  
(1996) which we reanalyze here, that this low luminosity elliptical has a convincingly unimodal distribution. However, our KMM tests suggest that  the color distribution of 
clusters in NGC 1427 is very likely to be bimodal. A recent ground-based 
analysis of this GCS in the Washington photometric system by  Forte et al. (2000) also finds a bimodal distribution. This gives us further confidence that
 the other low luminosity galaxies in our sample marked as likely bimodal candidates (NGC 1439, NGC 3377, NGC 4660; see note below on NGC 4486B) are indeed so. Thus, the merger model of cluster formation cannot be ruled out for low luminosity ellipticals. On the other hand bimodality in less massive galaxies poses a serious problem for the  C\^{o}t\'{e} et al.  (1998) model  of giant galaxy GCS formation for the reasons outlined below.

	In Fig 4 we plot the mean GCS colors of the galaxies with confirmed bimodal distributions and the ones with confirmed lack of bimodality as a function of host luminosity. Both distributions appear indistinguishable with
no evidence that the unimodal galaxies follow any kind of well defined  color(metallicity)-luminosity relation. To recall, the  C\^{o}t\'{e} et al.  (1998)  model assumes that the input low luminosity galaxies that are cannibalized to form giant ellipticals have a second order luminosity-cluster metallicity relationship. In fact the success of this model is critically dependent on the second order nature of the systems being cannibalized.  A set of merging/coalescing cluster systems with widely varying metallicity distributions, as seen in Fig 4, is  unlikely to produce a bimodal distribution in the absence of such a relation, which suggests that the  C\^{o}t\'{e} et al.  (1998) model is unviable. Moreover,  within the range of host luminosities of our sample we find no unimodal cluster system with a mean GCS color of V-I$\approx$1.2 (corresponding to the red peak of giant galaxies) which could serve as the progenitor host for the C\^{o}t\'{e} et al. cannibalism model. Since the upper envelope of the GC metallicity-host galaxy luminosity relationship appears to be defined quite sharply for both S0s and ellipticals (Fig 2), this suggests that if the  C\^{o}t\'{e} et al.  (1998) model is to be believed, the progenitor host galaxy
with the metal-rich cluster system must have an initial mass corresponding to a present day galaxy with an absolute magnitude of M$_T^V$$\approx$-23 mag. This is {\it more massive} than the typical
 M$_T^V$$\approx$-22.5 giant elliptical galaxy like M87 that the
model sets out to explain. We suspect that the reason for the shallower input 
color-luminosity relationship  used by C\^{o}t\'{e} et al.  (1998)  is that
they calculated it on the basis of ground-based data. It has been shown that in 
galaxies with bimodal color distributions the blue clusters are more spatially
extended than the red ones (e.g. Geisler, Lee \& Kim 1996; Kundu et al. 1999), so it is very likely that a relationship derived on the basis of ground-based data, that mostly samples the outer regions of a cluster system, returns a shallower color-luminosity relationship. The low luminosity elliptical galaxies present
a further problem for the C\^{o}t\'{e} et al model. In order to create bimodal low mass systems through the 'cannibalism' model one 
would require a population of low mass, unimodal metal-rich systems, which could then accrete other low mass metal-poor systems. There are two problems with this hypothesis: We observe no low mass metal-rich systems in our sample. And such
a requirement would violate the smooth second order metallicity(color)-host luminosity relationship required for the formation of bimodal systems in giant galaxies through the  C\^{o}t\'{e} et al model. 
Thus the observed bimodality
 in low luminosity galaxy could not have been caused by the mergers of two or
more systems of different metallicities via the 'cannibalism' model, and some other mechanism is at work. If
such an independent method for creating bimodality exists in low luminosity galaxies, one could turn this argument around and claim that the same process
could occur in giant galaxies. It may still be possible to attempt to retrieve 
the C\^{o}t\'{e} et al.  (1998) model for the central cD galaxies in large clusters  by postulating that the progenitor  galaxy is somehow inordinately
metal-enriched and cannibalizes primarily the metal-poor clusters of dwarf galaxies,
but such an explanation cannot explain bimodality in ellipticals that are not the dominant central galaxy in a cluster. While it is very likely that giant galaxies cannibalize dwarf galaxies and their metal-poor cluster systems to some extent, it is unlikely to be the primary mechanism of creating bimodality in a majority of galaxies. Thus, our observations do not seem to support the C\^{o}t\'{e} et al.  (1998) model and suggests that the bimodality in globular cluster colors is evidence for the formation of clusters in two different 
epochs of the metal-enrichment process. By extension this implies that for old cluster systems the red clusters are younger than the blue ones.

	There are no obvious ways to distinguish between the multiple collapse
and merger models based on the colors of clusters. They both predict bimodal color distributions with the red clusters situated closer to the center of the
 galaxy than the blue ones. While the merger model provides a mechanism for the second episode of cluster formation, one of the drawbacks of the multiple collapse model is that at present there is no known viable trigger. It is  possible that the second burst of cluster formation in the collapse model is 
triggered by a merger. One of the criticisms of the merger model is that, for it to be able to produce the thousands of metal-rich clusters seen in cD galaxies, it requires a huge gas reservoir ($\sim$10$^{11}$ Solar masses in the case of M87). This massive gas budget, coupled with the roughly similar ages of the red and blue clusters observed in galaxies like M87,
 suggests that for the merger model to work the event must have taken place very early in the history of the progenitors when they were still gaseous. The difference between the merger of two largely gaseous bodies and the collapse of
one large gaseous entity (with possible  fragments within this body) may largely
be one of semantics. 

	Various attempts have been made to correlate the properties of the present day galaxy with those of either the red, or the blue systems in order to
try and associate one population with the progenitor. These attempts are usually
 inconclusive, mostly because the present day galaxy might have significantly different morphological and/or chemical properties than the progenitor that formed the early clusters. However, Forbes et al.  (1997) and Forbes \& Forte (2000)  claim to find a better correlation between the host galaxy luminosity and velocity dispersion respectively (or mass) and the metallicity of the red clusters, as compared to the blue ones (The latter study is based in part on the results from this paper, previously published as part of the first author's PhD thesis). 

In Fig 5 we plot the peak V-I color of the blue and red peaks as a function of the absolute magnitude of the host. We have marked the peaks of the 
galaxies with the most obvious bimodality (see Table 3) with different symbols. 
The faint galaxy at M$_T^V$=-17.7 is NGC 4486B. Its cluster system is likely 
contaminated by the halo clusters of M87 so it may be somewhat dubious to plot the points at M$_T^V$=-17.7. The radial density profiles for M87 derived by McLaughlin, Harris \& Hanes (1994) yield a density of $\sim$8 clusters per square arcmin, or
$\sim$40 clusters within the WFPC2 field of view at the location of NGC 4486B. Given that the 7.3$'$ radial distance between NGC 4486B and M87 corresponds to the outermost bins of the 14$'$$\times$14$'$ field of view of the McLaughlin et al. analysis, and that the study was based on only one filter, this derived M87 cluster density is likely very uncertain. 

Fig 5 appears to agree with the Forbes et al. and
Forbes \& Forte  observation that the red clusters are better correlated with the luminosity than the blue ones, although the correlation is weak at best. In fact if we consider
  only the galaxies with the strongest evidence for bimodality the correlation essentially disappears. Even if we were to believe that the weak correlation 
of Fig 5 as real, it is neither very surprising, nor very informative, as it does not favor either the merger or the multiple collapse model; it suggests that while all
the hosts  have a blue population with a similar, most likely primordial metallicity, the metal-rich clusters are formed from gas that has been processed by the galaxy. Given that the stars in the more luminous galaxies also have larger 
metallicities it is not very remarkable that clusters that formed coevally from this enriched gas  also show such a trend. 

Finally, we would like to draw attention to the curious GCS of the elliptical galaxy NGC 7626, which
 appears to have a single metal-rich peak. A single metal-rich cluster component
 is incompatible with both the merger model and the multiple collapse model and
would present  a fairly significant mystery. A few other such systems, like IC 4051 (Woodworth \& Harris 2000) and NGC 3311, have been  identified in the literature, but doubts have been raised as to whether this is a real feature or
 an artifact of the photometric uncertainties (Woodworth \& Harris 2000; Brodie, Larsen, \& Kissler-Patig 2000). On closer examination of NGC 7626 we find that the Burstein \& Heiles (1982) reddening values used by us are substantially
smaller than the more recent Schlegel et al. (1998) numbers. Applying the 
Schlegel et al. reddening corrections shifts the mean V-I color of the NGC 7626 
GCS 0.05 mag blueward to 1.09 mag. Given the large reddening correction quoted by Schlegel et al. (1998), and the correspondingly larger absolute uncertainties induced by the extinction curve, we think that it is possible that the reddening is underestimated and that this galaxy does not in reality have a single metal-rich peak.  Furthermore, NGC 7626 is by far the most
distant galaxy in our sample. Deeper observations yielding more clusters may yet
show a bimodal cluster distribution in this galaxy. We also note that there are no luminous galaxies with a single metal-poor peak in our sample. The NGC 4874 system discovered by Harris et al. (2000) is unique in this aspect.

	Based on our sample of 29 elliptical galaxies - including M87 - we conclude that a majority of the globular cluster systems 
of elliptical galaxies are formed in two episodes. The second episode is most
likely triggered by a merger; possibly a major merger in the case of cD galaxies
like M87 and a minor merger in the case of S0 galaxies like NGC 3115 that
preserves the disk component (Kundu \& Whitmore 1998).

\subsection{The Globular Cluster Luminosity Function as a Distance Indicator}

	The turnover luminosity of the GCLF has been found to be remarkably constant over a wide range of galaxies and environments. So constant in fact that it has been used as a secondary distance indicator (Harris 1991, Jacoby 1992; Whitmore 1996). The theoretical basis for this rather remarkable
result - which implies that for any reasonable range of M/L ratio the underlying
mass distribution of globular clusters is the same in all galaxies - is not well
understood. There are lingering doubts that the GCLF might be affected by  small
 metallicity variations, which must be accounted for to get optimal usage as a distance indicator. In previous analyses (Whitmore et al. 1995; Kundu \& Whitmore 1998; Kundu et al. 1999) we showed that the GCLF  turnover magnitude gives reasonable estimates of the distance to program galaxies.  In this section we shall attempt to better calibrate the GCLF with known distance indicators and to comment on its usefulness as a distance indicator.
 
 	For an accurate determination of the globular cluster luminosity function it is imperative that we measure the completeness curve (i.e. the percentage of clusters detected per unit magnitude) and apply a correction term for this. The analysis of this sample of galaxies is somewhat complicated by the fact that the data set is not homogeneous.  The 
different exposure times, gains, galaxy backgrounds, and cluster color distributions require us to calculate the completeness curves for each galaxy individually. A detailed explanation of the procedure adopted by us can be found in Appendix A.

	Conventionally, the turnover luminosity of the GCLF is determined by
fitting a Gaussian curve, although the choice of a  Gaussian is rather ad hoc and
has no physical basis. While a Gaussian  seems to fit the GCLF quite well,
Secker $\&$ Harris (1993) suggest that a t$_5$ distribution provides a better
fit in many cases.  We correlated the Gaussian turnovers calculated by the method outlined below with  the turnover luminosity from a  t$_5$ 
distribution fit  using the maximum likelihood code of Secker $\&$ Harris (1993)
and found that they compared quite favorably. Moreover, we argue later in this section  that one can only accurately determine the turnover luminosity in systems where the faintest clusters allowed in the fit (the 50\% completeness
 limit) are brighter than the turnover luminosity, and the width of the fitting 
function is held constant. In such cases the choice of the exact fitting function has little effect on the turnover luminosity.

	The dispersion, $\sigma$, of the Gaussian describing the GCLF has been
reported as varying considerably from galaxy to galaxy in the literature. In order to test whether this is a real effect, or simply due to the varying quality of data used by various authors, we first selected a subsample of  11
galaxies with the deepest photometric data in the V and I-bands (as compared to the distance to the 
galaxy from Table 1) in our sample. For these 11 galaxies we fit  Gaussian curves to the 
 completeness corrected GCLF in both the V and I-bands using the IRAF task 
NGAUSSFIT. We allowed the algorithm to fit both the $\sigma$ and the turnover magnitude. For each GCLF we varied the width (0.18 to 0.25 mag) and positioning of the bins, and the cutoff 
magnitude at the faint end (40$\%$ to 55$\%$ completeness) and averaged the results. In all, we performed eight individual fits for each luminosity function
 and average the results to get a measure of the  turnover magnitudes, the associated Gaussian dispersions, and the errors in the mean of the eight fits.
 These numbers are listed in Table 4 (the galaxies are sorted by distance). The uncertainties are a  indicator of the effects of binning, the choice of a Gaussian, the number of clusters used for the fit, and the position of the 50\% cutoff limit
with respect to the turnover magnitude. By and large, the peak of the cluster distribution migrates
to fainter magnitudes for more distant galaxies. 

	In the top panel of Fig 6 we plot the dispersion in the V-band GCLF fit,
$\sigma$$_V$, vs that in the I-band, $\sigma$$_I$. In general the dispersion in the two bands seem 
to be correlated, although there is plenty of scatter in the relationship. In
 the middle panels we plot $\sigma$$_V$ and $\sigma$$_I$ as a function of the
mean colors of the GCS while in the bottom panels we plot the dispersions as a
function of the absolute magnitude (mass) of the host galaxy. In most cases we
note that the low metallicity (bluer) cluster systems in less massive galaxies have a
fairly constant dispersion 1.2-1.3 mag in both V and I. The metal-rich cluster
systems in more massive galaxies appear to have wider GCLFs. Since the very metal-rich systems in our sample all have bimodal distributions, and the luminosity functions of each of red and blue systems might be slightly offset 
(e.g. in M87, Kundu et al. 1999), it is not very surprising that the integrated
luminosity function of the entire system appears wider. Based on Fig 8 we  
conclude that GCSs of low mass galaxies and/or metal-poor systems
have a fairly constant GCLF dispersion of $\sigma$$\approx$1.25, while metal-rich GCSs have wider GCLFs. The mean dispersion of the GCLFs of the 11
galaxies is also listed in Table 4. We note that the numbers are rather similar 
in the two filters $\sigma$$_V$=1.32$\pm$0.15 (0.04) and $\sigma$$_I$=1.30$\pm$0.19 (0.06).

As is evident from Table 4 and Fig 6 the dispersions in the I-band are more 
scattered than in the V. We suspect that this is possibly due to the fact that, in general, our I-band observations are less deep, and more importantly the galaxy background is
stronger in I. This is borne out by the fact that the mean uncertainties in the
I-band magnitudes of individual clusters appear to be larger than those in the V for every galaxy in our sample, {\it even in cases where the I-band images are deeper}. Although the uncertainties in individual fits of $\sigma$$_V$ and $\sigma$$_I$ in Table 4  appear to be similar ($<$$\sigma$$_V$$>$=0.06 and $<$$\sigma$$_I$$>$=0.07), the uncertainty in the mean $\sigma$, signifying the scatter, {\it is} higher in 
the I-band (0.19 vs 0.15). This leads us to the intuitively obvious conclusion 
that it is possible to get excellent two parameter fits with low statistical
errors, although the actual systematic errors in each of the quantities is higher. We believe that the large scatter in $\sigma$ values reported in the literature is largely due to the attempts to fit two parameters to the
GCLF simultaneously on poor quality data. As is evident from our analysis, the GCLF dispersion and turnover can simultaneously be fit only for the deeper subsample of our HST images.

	For further analysis of the GCLF turnover we  fit the luminosity function of each GCS to a Gaussian of fixed width. While the mean $\sigma$ of the Gaussian fits in Table 4 is 1.3 in both the 
V and I-bands, it is also clear that the dispersions of the most metal-rich
 systems is slightly higher. In order to prevent any bias due to the choice of
the dispersion we fit the GCLF to Gaussians with fixed values of $\sigma$=1.1,
 1.2, 1.3, 1.4, 1.5, varying the position and size of the bins and the cutoff
magnitude at the faint end. The 
mean $\sigma$ for these 'fixed width' GCLF fits is 1.3. We report
the average of 40 such independent fits to the turnover luminosity of each GCLF, and the associated
standard deviation in the mean of the these values in columns 5 and 9 of Table 4. The
advantage of such averaged  'fixed width' fits is that they are not biased by
the exact choice of the fitting parameters and the error estimates are realistic
 indicators of the actual uncertainties in the turnover magnitude, including the
effect of the choice of a Gaussian as a fitting parameter. It is evident from
Table 4 that the turnover magnitudes and uncertainties from the 'fixed width' analysis are very similar to those from the 'variable width' fits , giving us 
confidence in the former numbers. In fact, in cases where the uncertainty in 
 $\sigma$ is large for the 'variable width' cases the turnover luminosity of
the 'fixed width' fits appears to be more secure. Although we have listed the
turnover luminosities of all 29 galaxies in our sample as an academic exercise,
 we would like to point out that those with an uncertainty greater than 0.3 mag are
likely to be highly unreliable since in most cases the images are not deep enough to
sample a significant fraction of the cluster system of these galaxies.  The turnovers with calculated uncertainties $>$= 0.3 mags should not be
 used in any serious analysis of GCLF turnovers. In the next section we integrate the total number of clusters under a Gaussian GCLF to calculate the
specific frequency. The calculated turnover luminosities and associated errors for the most distant galaxies are listed primarily to enable the reader to judge the effect of GCLF fitting on the specific frequencies of these hosts.

	In Fig 7 we plot the completeness corrected GCLFs for the 11 best determined galaxies in our sample and the best fitting 'fixed width', $\sigma$=1.3, Gaussian fits, in both the V and I-bands. The galaxies are sorted
by the distance from Table 1. As is evident from the figure, Gaussians provide an excellent fit to the GCLF and the apparent magnitude of the GCLF turnover migrates to fainter numbers
for more distant galaxies. We can further check for the internal consistency of the GCLF, and its utility as a distance indicator by comparing the
turnovers in V and I. 	In Fig 8 we plot the turnover luminosity in the V-band vs that in the I-band. It is immediately apparent that m$_V^0$ and m$_I^0$ are very tightly correlated. The uncertainties are larger for the fainter systems as is only to be expected from the completeness limits.

Ashman, Conti \& Zepf (1995) showed that if the mass function of GCs is
universal, the position of the peak of the GCLF is slightly dependent on the 
metallicity; for the V and I-bands it shifts to fainter magnitudes for more metal-rich systems, the effect being larger in V.  In Fig 9 we plot the variation in m$_V^0$-m$_I^0$  vs the mean metallicities from Table 3 for candidates with $\delta$(m$_V^0$-m$_I^0$) $<$ 0.15 mag.  We have also included
 M87 (Kundu et al. 1999) in the plots.  It is evident that for both the 'variable width' fits and the 'constant width' fits the difference between the
 turnovers increases with metallicity in a manner consistent with the Ashman et
 al. (1995) prediction. The amplitude of this variation is also consistent with 
the Ashman et al. values.

But how well do these turnover luminosities  trace the distance to the galaxy? 
While the recent compilations of Kavelaars et al. (2000) and Whitmore (1996) 
show that the GCLF is an excellent distance indicator Ferrarese et al. (2000), in their comparison of various distance indicators, suggest that the utility of
the GCLF as a distance indicator is questionable. In order to test the consistency we compare the GCLFs with the weighted distance moduli of individual
 galaxies calculated by Ferrarese et al and the SBF distances measured by Neilsen (1999). We note that in most cases Neilsen uses
the  same HST as we do. The distances measured by these two groups,
and the Table 1 distances are listed in Table 6 (columns 2 - 4).

	To calculate the absolute magnitude of the turnover luminosity we compared the GCLFs with the best determined turnovers i.e. uncertainty less than
0.1 mag, with the weighted distance moduli of three other distance indicators from Ferrarese et al. (2000) and Neilsen
 (1999). The turnover magnitudes for both the 'fixed width' and variable width
cases along with the standard deviation and number of galaxies used in calculating the difference are listed in the upper half of Table 5. It is evident that the GCLF turnover is in excellent agreement with the distance measurements using other methods, and that the 'fixed width' 
Gaussians are more accurate than the 'variable width' ones.   In comparing the Neilsen values with the
Ferrarese et al. numbers (again restricting the Neilsen sample to those with an uncertainty of less 0.1 mag for the sake of consistency) we find that the uncertainty in the difference, 0.14 mag, is comparable with the GCLF values (0.11 and 0.14). Thus the GCLF is {\it as accurate a distance indicator as the SBF}. In fact, as the weighted Ferrarese et al. distances include SBF
 measurements and do not include GCLF measured distance moduli, the comparable uncertainties in Table 5 suggest that the GCLF method may even be slightly  superior to SBF (based on a small sample of 5 galaxies for the SBF comparison). 
In the lower half of Table 5 we report the uncertainty weighted mean difference
between the 'fixed width' turnovers and the Ferrarese et al. and Neilsen distance measurements.  Since the Ferrarese et al. distances are
based on a weighted mean of various distance indicators, the 'fixed width'
 turnover comparison with these numbers are likely to be most accurate.
Thus we calculate a turnover luminosity of M$_V^0$ = -7.41(0.03) in V, and 
M$_I^0$=-8.46 (0.03) in the I-band.

In Fig 10 we compare the GCLF turnover luminosities, in both the V and I-bands, with the Ferrarese et al. mean distance moduli and the Neilsen SBF distances. We also compare the Neilsen distances to
the weighted means. The GCLF turnover tracks the distance to galaxies
 excellently throughout the entire distance range and appears slightly more reliable than the SBF method.

	In Table 6 we list the GCLF distances
using M$_V^0$ = -7.41 and M$_I^0$=-8.46. The distances derived from the V and I-bands appear to be in excellent agreement with each other. We have only listed
candidates with mean GCLF distance modulus uncertainties less than 0.2 mag which we consider to be reliable. Ashman et al. (1995) suggest that the metallicity dependent effect on the GCLF (Fig 9) should largely be applicable to the V-band turnovers. However, on correcting the GCLFs of our
 sample using the relationships derived in Fig 9 and comparing with the Ferrarese at al. and Neilsen et al. distances we find no obvious improvement in the uncertainty of the turnover. It may be that  better quality data is required
 to test for such small corrections of less that 0.1 mag.

	We have shown above that the GCLF turnover is in excellent agreement 
with the Ferrarese et al. (2000) distances. Why then do Ferrarese et al. claim
 that the GCLF is not a reliable distance indicator? The answer may lie in the comparison sample used by them. Ferrarese et al. base their conclusion mainly on
 3 data points in the V-band. While two of the points correspond to the 
weighted mean of turnovers in two sub-clusters of the Virgo cluster the third 
point is based on the mean of 4 galaxies in Fornax. The two points in Virgo are
 based on deep HST-data and are in excellent agreement with each other, and 
other distance measurements. On the other hand, the weighted mean of the Fornax turnovers  is entirely based on old ground-based data. The thrust of the criticism of the GCLF 
method by Ferrarese et al is largely that the mean distance modulus of 
these four Fornax galaxies is similar to the members of the Virgo cluster and
not 0.5 mag fainter as suggested by the Cepheid distances. 
As we have 
shown earlier in this section it is only possible to accurately determine the
GCLF turnover for a few of the deep HST images at this distance. It is very
likely that the old ground-based data for these Fornax galaxies is not sufficient to accurately determine the turnover. We should also point
 out that incomplete GCLF data tend to return a systematically lower value of 
turnover (Secker and Harris 1993), which could easily explain this discrepancy.
Furthermore, one of the V-band GCLFs for NGC 1399 used by Ferrarese et al. (Ostrov, Forte \& Geisler 1998) is actually based on Washington photometry, with
all the uncertainties associated with filter system transformations. The only
deep HST-based study of Fornax GCSs is by Grillmair et al. (1999). Unfortunately,
they  report their turnover magnitudes in the B-band which precludes us from
making a direct comparison with our data. However, we note that NGC 1427, the 
only Fornax galaxy in our sample, has a turnover luminosity that is 0.5 mags fainter than the members of the Virgo cluster. While the uncertainties associated with the NGC 1427 turnover are rather on the high side it is in fact consistent with the distance modulus using other methods reported by Ferrarese
et al.

	Even if one were to assume that the ground-based GCLF turnovers in Fornax are accurate, the GCLF distances are far less discordant than made out by Ferrarese et al. From their Table 2 we compute the mean distance moduli and dispersion for each of the four
distance indicators in Fornax and Virgo (M87 + NGC 4472 Groups). The difference in the two groups i.e. (m-M)$_{Fornax}$ - (m-M)$_{Virgo}$ for each of the methods is: 

Cepheids: +0.20

Planetary Nebula Luminosity Function: -0.01

V-band Globular Cluster Luminosity Function: -0.28

I-band Surface Brightness Fluctuation: +0.09

	Ferrarese et al. (2000) assume the Cepheids  as a standard  for  comparisons of Fornax vs. Virgo measurements. However, the Cepheids appear nearly as much of an outlier as the ground-based GCLF measurements, and their conclusions are biased by adopting the Cepheids as the benchmarks for measuring the accuracy of other distance indicators. Furthermore, the unweighted mean Cepheid distance to the Fornax cluster is m-M = 31.54$\pm$0.23 ( (m-M)$_{NGC1326A}$=31.43$\pm$0.07; (m-M)$_{NGC1365}$=31.39$\pm$0.10; (m-M)$_{NGC1425}$=31.81$\pm$0.06 ). The scatter in the Cepheid distance measurements  in Fornax is much larger than the internal precision of the method and it is not obvious that the Cepheid values can be used as a standard candle to test the viability of the GCLF method (or any of the other methods for that matter).

 In our view Fig 10 and Table 6 show rather convincingly that the GCLF method is
 in excellent agreement with other distance indicators, including those of Ferrarese et al. Thus we conclude, on the basis of the study of the GCLFs of our sample galaxies, that the turnover luminosity is an excellent distance indicator with 
an accuracy that is comparable, and may even be superior to the SBF method.

\subsection{Specific Frequency}

	The specific frequency of a cluster system   was introduced by Harris \& van den Bergh (1981) as a measure of the relative efficiency of cluster formation in galaxies (as compared to stars) and is defined as: S$_N$ = N $\times$ 10$^{0.4(M_V + 15)}$. For the small field of view of our HST images we can only calculate the local
specific frequency. In order to do so we need to determine the projected total number of cluster candidates in our images.
While calculating the GCLF turnovers for Table 4 we also measured the amplitude
of the best fit Gaussian for the 'fixed width' fits. We calculated the mean
amplitude of the V and I-band GCLFs for each galaxy (after weighting the data
appropriately to account for the variable bin sizes in the 40 individual fits),
and then integrated  the area under this $\sigma$=1.3 Gaussian curve to calculate the total projected number of clusters in the field of view.  In order to calculate the luminosity of the galaxy in our field of view we measured the total integrated light within the F555W image and subtracted the "typical" sky background of 0.052 e$^-$s$^{-1}$pixel$^{-1}$ for the WF and 0.010 e$^-$s$^{-1}$pixel$^{-1}$ for the PC. Since the host galaxies cover the entire WFPC2 field of view in each
 of our images it is impossible to measure the background from the images.  The "typical" sky background level is thus based on the estimates in the HST Data Handbook (1997). Furthermore we estimated the error in the surface photometry to be of the order of this background value and propagated this in the calculation of the uncertainty in the local specific frequency. 

In order to further test the accuracy of the surface photometry, we compared   V-band magnitudes in small apertures with published aperture photometry. For an aperture of 30.1$''$ in NGC 4649 we derived V = 10.86 mag, compared to the published value of 10.80$\pm$0.02 mag (Sandage \& Visvanathan 1978). For apertures of 25.9$''$ and 29.8$''$  in NGC 1427 we calculated V = 12.29 mag,  and 12.18 mag, in good agreement with the published values of 12.35$\pm$0.02 mag and 12.19$\pm$0.02 mag respectively (Sandage \& Visvanathan 1978; Persson, Frogel \& Aaronson 1979). The published  V = 11.20$\pm$0.02 mag (Sandage \& Visvanathan 1978) for a 30.1$''$ aperture is consistent with our values of 11.19 in NGC 4552. Similarly in NGC 3379 the reported magnitudes of 11.21$\pm$0.06 and 10.9$\pm$0.06 for 20.9$''$ and 29.0$''$ apertures are in excellent agreement with our numbers of 11.14 mag and 10.9 mag. Thus the aperture photometry, and hence the specific frequencies derived in this section are very reliable within the error budget.

	The projected total number of clusters and the local specific frequencies of our sample galaxies are listed in Table 7. For the most distant galaxies in our sample only a small fraction of the clusters brighter than the GCLF turnover are observed. As the cluster population of these distant hosts are based on the extrapolation of the GCLF past the turnover, the specific frequencies of these candidates naturally have very large uncertainties. 
The local specific frequencies derived in this analysis and in Paper II are of course measured within the WFPC2 field of view, hence a larger fraction of the cluster population and galaxy light is included for
the more distant galaxies. While specific comparisons between individual galaxies in either sample may not yield very meaningful results, on average the ellipticals in this paper and the S0s in Paper II are distributed over a similar
  range of distances [(m-M) = 31.4$\pm$0.7 for ellipticals and (m-M) = 31.2$\pm$1 for S0s]. Hence we can legitimately make comparisons between the average properties of both samples. 
Also, in consort with published values, it is possible to study the radial properties of the specific frequency of each galaxy. The local specific frequencies calculated here add to this reference list.
The mean local specific frequency of our elliptical sample, for systems with $\delta$S$_N$$<$3 is 2.4$\pm$1.8 (0.4). This is significantly larger than the local specific frequency of the S0 sample in Paper II, S$_{N(local)} = $1.0$\pm$0.6 (0.1), and consistent with previous observations that the specific frequency of ellipticals in on average 2-3 times larger than than that of S0s (Harris 1991). 
 
	Fig 11 shows the variation of local specific frequency with mean GCS color (metallicity), the luminosity (mass) of the host galaxy, and the number
of galaxies in the host (galaxy) cluster (Garcia 1993).    There appears to be no obvious correlations between the specific frequency and any of the  quantities. Previous studies of the properties of GCSs of elliptical galaxies (Ashman \& Zepf 1998) have
reached similar conclusions. While the Ashman \& Zepf (1992) merger model predicts that the specific frequency of metal-rich cluster systems must be higher than that of metal-poor systems due to the addition of the younger
metal-red population, this is probably a somewhat
simplistic view since large ellipticals usually have significant populations
of blue clusters which may have been cannibalized from nearby dwarfs. The lack
of a correlation in Fig 11 suggests that no one single process determines the 
specific frequency of globular cluster systems. In order  to better understand the underlying reason for the difference in specific frequency (or lack thereof)
from galaxy to galaxy, and to further discriminate between the various formation scenarios it is imperative to study the global specific frequency of the cluster systems and the spatial distributions of the metal-rich and metal-poor subsystems. While Forbes et al. (1997) attempted such an analysis based on early
observations of a handful of galaxies, a recent detailed multicolor study of the NGC 4472 system by Rhode \& Zepf (2001) suggests that the previous studies underestimated the effects of contamination. Rhode \& Zepf report a specific frequency of 3.6$\pm$0.6 within r=23$\arcmin$ as opposed to the values of S$_N$
between 5 and 7 reported in the literature (Lee, Kim \& Geisler 1998; Harris 1986; Harris 1991).
More such detailed global studies will go a long way towards sorting out the formation scenarios. 

	We also note that in each case where the global specific frequency has
been calculated from ground-based data it appears to be larger than the local specific frequency calculated by us (Table 7). This suggests that S$_N$ is lower in the
innermost regions of the galaxies that the WFPC2 images sample and is consistent
with previous observations that the globular cluster density function in the
innermost regions of galaxies is shallower that the galaxy light profile (Forbes et al. 1996) 

\subsection{Cluster Sizes: Another Distance Indicator?}

	 In all the clusters systems that we have studied with  HST, we have found the PSF to be slightly broader than that of an unresolved star. This sample of elliptical galaxies is no exception.  Using the technique described in Kundu \& Whitmore (1998) we measured the sizes of individual candidates by comparing the profiles of the clusters 
 to those of PSF convolved King models. We restricted our analysis to only the
cluster candidates on the PC chip since the undersampled PSF in the WF
 introduces excess uncertainties in the size determination. Furthermore, we
restricted our input library of profiles to  King models with a fixed parameter
 c=1.25. 

 Using the distance moduli from Table 1, we then converted the angular sizes of individual clusters to linear distances. Fig 12  is a plot of the linear 
half light radius of individual clusters vs the V mag. It is immediately apparent that the sizes of a majority of the clusters fall in a narrow range between 0.5 to 4 pc. It also appears that in every nearby galaxy (Table 1) with a deep image (Table 2) the cluster sizes peak at a half light radius of $\approx$2.5 parsecs and that there is no correlation of the sizes with luminosity. This is a rather interesting discovery which suggests that globular clusters have a preferred size. There is
evidence that there is a similar peak in the size distribution of Galactic globular clusters at about 3 pc (van den Bergh 1996). In Table 8 we list the median sizes of the 
cluster candidates in the PC in parsecs and the number of objects in the chip. 
Apart from a few outliers the median size appears to be remarkably constant. The mean
value for our entire sample is 2.36$\pm$0.4 (0.08) pc. 

 To further illustrate this point we have plotted the median sizes of the 
cluster system in a given galaxy as a function of distance in the top panel of Fig 13. There appears to be a  trend of increasing median
size with distance to the galaxy. This is most likely an artifact of using a model PSF which may not perfectly fit the observed profile of a point source e.g. if the model PSF is too sharp / centrally peaked. Coupled with the undersampling of the WFPC2 such an error would induce just 
the kind of distance dependent increase in the measured sizes seen in Fig 13. Consider cluster candidates in two galaxies, one nearby and another distant. In the former case  the measured width of the cluster profile will largely be governed  by the actual physical radius of the cluster (in other words the cluster is well resolved), while for the more distant galaxy the profile is dominated by the instrumental PSF (the slightly resolved clusters increase the width by a few percent). If the adopted PSF is too centrally peaked - imagine a
delta function in the extreme case - the King model convolved PSF will appear only slightly narrower than the actual profile for a nearby galaxy, while it will appear considerably narrower for a more distant host. Thus, in attempting to fit the model profile to real clusters one would overestimate the size of GCs
 in the nearby galaxy by a small amount	and those in distant ones by a larger margin, thereby leading to the trend in Fig 13. Even though we have modeled the cluster profiles taking into account
 many of the idiosyncrasies of the WFPC2 PSF, the largest source of error is still in the adopted model PSF. A more accurate PSF may reduce the scatter in the sizes even further.

One could argue, based on the fact that we detected our cluster candidates using fixed concentration criteria for all the galaxies in $\S$2, that we are selectively missing the largest clusters in the nearby galaxies, thereby inducing this distance dependent size variation. However, numerical tests suggest that we should easily detect clusters with r$_h$ greater than 15 pc even in the nearest candidates. It is evident from Fig 12 that the largest clusters measured in 
nearby galaxies e.g. NGC 4472, NGC 4649 and NGC 4406 are of the same  size - 
$\sim$6 pc - as the largest candidates in the most distant galaxy in our sample, NGC 7626 which suggests that there are few, if any, clusters with half light radii greater than 6 pc in the inner regions of galaxies. It is also known that the median size of Galactic globular clusters
increases with Galactocentric distance. One might further argue that the larger
median size of the the more distant systems is due to a larger degree of contaminations by such large clusters. However, in comparing the sizes within
the much larger effective areas covered in our entire WFPC2 images  of M87  and NGC 3115 (Kundu et al. 1999; Kundu \& Whitmore 1998), covering a range of galactocentric radii from R = 0 to R = 8 kpc, we conclude that this is not likely to be a major effect for the clusters in the PC chip considered here. Thus, we are quite convinced that the distance dependent effect is a 
consequence of the PSF.

In the middle panel of Fig 13 we plot only those galaxies that have at least 30 clusters in the PC. The better statistics of this sub-sample noticeably reduces the variation in median cluster sizes from galaxy to galaxy with a mean value
of  2.25$\pm$0.2 (0.07) pc.  One drawback of using the Table 1 distances 
is apparent in the middle panel of Fig 13; as the same distance modulus is used 
for all the galaxies in a particular cluster (Virgo) the physical depth of the cluster is translated into additional scatter in the calculated cluster sizes. To get around this problem we plot
 the distances and the median spatial sizes of the same set of galaxies as in 
the middle panel using the mean GCLF distances from Table 6, in the bottom panel of Fig 13 (for candidates with uncertainties in GCLF distance moduli less than 0.2 mag). The linear sizes have also been recalculated using the GCLF distances. Again we see a small distance dependent effect that we suspect is
 due to the PSF. In the absence of this effect the median cluster sizes in the 
inner regions of our candidate galaxies may have been even more constant than it
 appears from our calculations. The mean size of the Virgo sample  is 2.2$\pm$0.2 pc using the distance estimates from Table 1 and 2.4$\pm$0.3 pc using the GCLF distances. Since the calculation of the median sizes involves
the distance estimate to a particular galaxy it may be more instructive to compare the actual measure quantity, the mean size in arcsecs; the mean size of the clusters in the Virgo cluster is 2.8$\times$10$^{-2}$ $\pm$ 2.6$\times$10$^{-3}$ arcsecs. We estimate that the inherent depth of $\sim$2 Mpc of the Virgo cluster, the PSF induced effects, and the 0.1 - 0.2 mag accuracy of the distance moduli  to individual galaxies would contribute $\sim$10\% uncertainty in the mean size of the Virgo sample. Thus, based on the 10$\%$ accuracy of the cluster sizes in Virgo it appears likely that in the absence of instrumental effects the median sizes of clusters is likely to be constant.

	If, as appears from our data, the median cluster size is indeed same in all galaxies it provides an exciting new way to measure the distance to a galaxy by purely geometrical considerations. Dynamical models (Murray \& Lin 1992 and references therein) show that the half light radius is a remarkably robust quantity that is 
largely unaffected by changes in the core of the clusters or the tidal truncation in the outer region; thus the half light radius is quite immune
to  tidal forces. A major advantage of  using $r_h$ as a distance indicator is that 
we only need to observe the brightest part of the luminosity function in order to determine the size; the scatter in sizes is smallest for the bright objects and there is no magnitude dependence of the average size. This is evident from the measured r$_h$ of clusters in the Milky Way (van den Bergh 1996), M87 (Kundu et al 1999), and some of the richer nearby cluster systems in our sample (Fig 12). The only limiting factor of this method is  the angular resolution of the telescope. 
We estimate that we can measure distances out to $\sim$70 Mpc with  the next generation Advanced Camera for Surveys (ACS) which will provide well sampled PSFs over a much larger field of view.

\section{Summary}

	We have analyzed the  globular cluster systems of  28 elliptical galaxies from archival WFPC2 images. In each galaxy we detected a population of old  globular clusters with colors
 in the range 0.5$<$V-I$<$1.5 mag.  
 
We find that at least 50$\%$ of the globular cluster systems have bimodal, V-I  color distributions {\it at the present level of photometric accuracy}. The mean V-I color of the GCS of these 28 galaxies is 1.04$\pm$0.04 (0.01) mag, corresponding to a mean metallicity of -1.0$\pm$0.10 (0.05).  

In this analysis we also  present the first evidence of possible bimodality in the globular cluster color distributions of low luminosity elliptical galaxies NGC 1439, NGC 1427, NGC 3377 and NGC 4660. Based on our data, the C\^{o}t\'{e} et al.  (1998) model of
multimodal color distributions by aggrandizement appears unlikely to be the primary mechanism for the formation of globular cluster systems of giant elliptical galaxies. It appears that the bimodal color distributions are a result of multiple epochs of globular cluster formation in
the metal enrichment history of the galaxy. The second event of cluster formation may have been triggered  either during a major merger event or in multiple collapse modes.

 The globular cluster luminosity function is an excellent distance indicator with an accuracy that is as good as, or probably better than the surface brightness fluctuation method. We find evidence that the difference in the turnover luminosities in V and I increases with metallicity, as predicted by Ashman et al.  (1995). We calculate
a turnover luminosity of M$_I^0$=-8.46 (0.03) in the I-band and M$_V^0$=-7.41 (0.03)  in the V-band.
 
The mean local specific frequency of the elliptical galaxies in our sample is 2.4$\pm$1.8 (0.4) which is significantly higher than that of our S0 sample (Paper II  1.0$\pm$1.1 (0.2)). There are no obvious trends between the local specific frequency and the mean metallicity if the cluster systems, the host galaxy luminosity, or membership of the galaxy in a galaxy cluster.

 The median size of the cluster candidates appears to be constant in all galaxies. We propose that this may be used as a simple geometrical means of determining the distances to galaxies in the future.

	AK is grateful to Mike A' Hearn, Francois Schweizer, Sylvain Veilleux 
and Stuart Vogel for numerous suggestions that helped improve this paper in its 
earlier incarnation as a chapter of his thesis. Yan Fernandez for all his 
suggestions and help. The authors would also like to thank the anonymous referee for the many useful comments and suggestions. Support for this work was provided by NASA through grant number AR-8378 from the Space Telescope Science Institute,
 which is operated by AURA, Inc., under NASA contract NAS5-26555. This research 
has made use of the NASA/IPAC Extragalactic Database (NED) which is operated by 
the Jet Propulsion Laboratory, California Institute of Technology, under
 contract with the National Aeronautics and Space Administration.

\appendix
\section{Completeness Correction}

 In order to calculate the completeness curve for each candidate galaxy we first created  PSFs for each chip and filter from the cluster candidates using the IRAF task PSF within the DAOPHOT package. We then added 100  simulated clusters per chip per simulation in the same location on the V and I image, with a color distribution similar to that of the globular
clusters in the particular galaxy. Next we attempted to detect the model GCs using our detection routine, with the exact same selection criteria as in \S 2. In all we simulated $\approx$20000 objects between 19$<$V$<$26 in each galaxy. The I-band simulated cluster magnitudes were typically 1 mag brighter though the exact input distribution varied from galaxy to galaxy because of the color range enforced. We then computed  the completeness curves for various background levels, in each filter, in both the PC and the WF, for every galaxy in our sample.
 
 Ideally one would prefer to
 divide the GCLF into smaller sub-samples in regions where the completeness is
reasonably similar and correct the GCLF in each of regions by the curve that describes it. The problem with this approach is that few galaxies have large enough samples of clusters for us to be able to apply the completeness corrections in eight to ten different regions, as is required for accurate  correction. Further subdivision makes statistical corrections meaningless. For example, if we
were analyzing the GCLFs in a fairly bright galaxy with a bright nucleus centered on the PC, it is entirely possible that we detect no clusters fainter than V$\approx$23 mag in the innermost region while the 50$\%$ completeness limit in the WF is at V$\approx$24 mag. In such a scenario we could not correct the inner GCLF past
 V$\approx$23, or carry out any legitimate analysis of the entire GCLF past this magnitude without eliminating the inner data points. To get around this problem we opted, instead, to calculate the weighted completeness curve for each galaxy and correcting the entire GCLF using the following method.  We first selected a  fiducial magnitude ($\approx$23 mag for the V-band GCLF and $\approx$22 mag for the I-band in most cases)  such that the
total number of clusters brighter than this magnitude is largely unaffected by
 completeness effects even in the regions of high background counts. We then 
weighted the completeness curves at various background levels by the relative 
number of bright objects in that region to arrive at the final completeness
 curve. One could argue that in calculating this weighted curve
 we are implicitly assuming that the GCLF has a similar shape everywhere (within a galaxy). 
Since the peak of the GCLF in no galaxy observed to date has ever been 
conclusively proven to vary by more than $\approx$0.2 mag even in the innermost 
region this appears to be an acceptable premise for the purpose of calculating 
the completeness curve. Besides, this correction is a small second order effect that affect the GCLF at less than a $\sim$0.1 mag level. We should also note that in calculating the completeness curves we have deliberately chosen
 to detect clusters in both the V and the I-band simultaneously in order to 
compare with our final color selected cluster sample. We could have chosen to 
separately detect clusters and calculate completeness curves for each filter to 
 reach a fainter detection limit (as the completeness limit would 
not be affected by the fainter of the two filters). However, we believe that the
gain in eliminating contaminating foreground/background objects from the color 
selected sample far outweighs the advantage of reaching a fainter detection limit in one filter.  The 50$\%$ 
completeness limits for each galaxy in the V and I filters are noted in Table 4.

\newpage
\renewcommand{\baselinestretch}{1}

\phantom{a}
\begin{figure}[!ht]
\centerline{\psfig{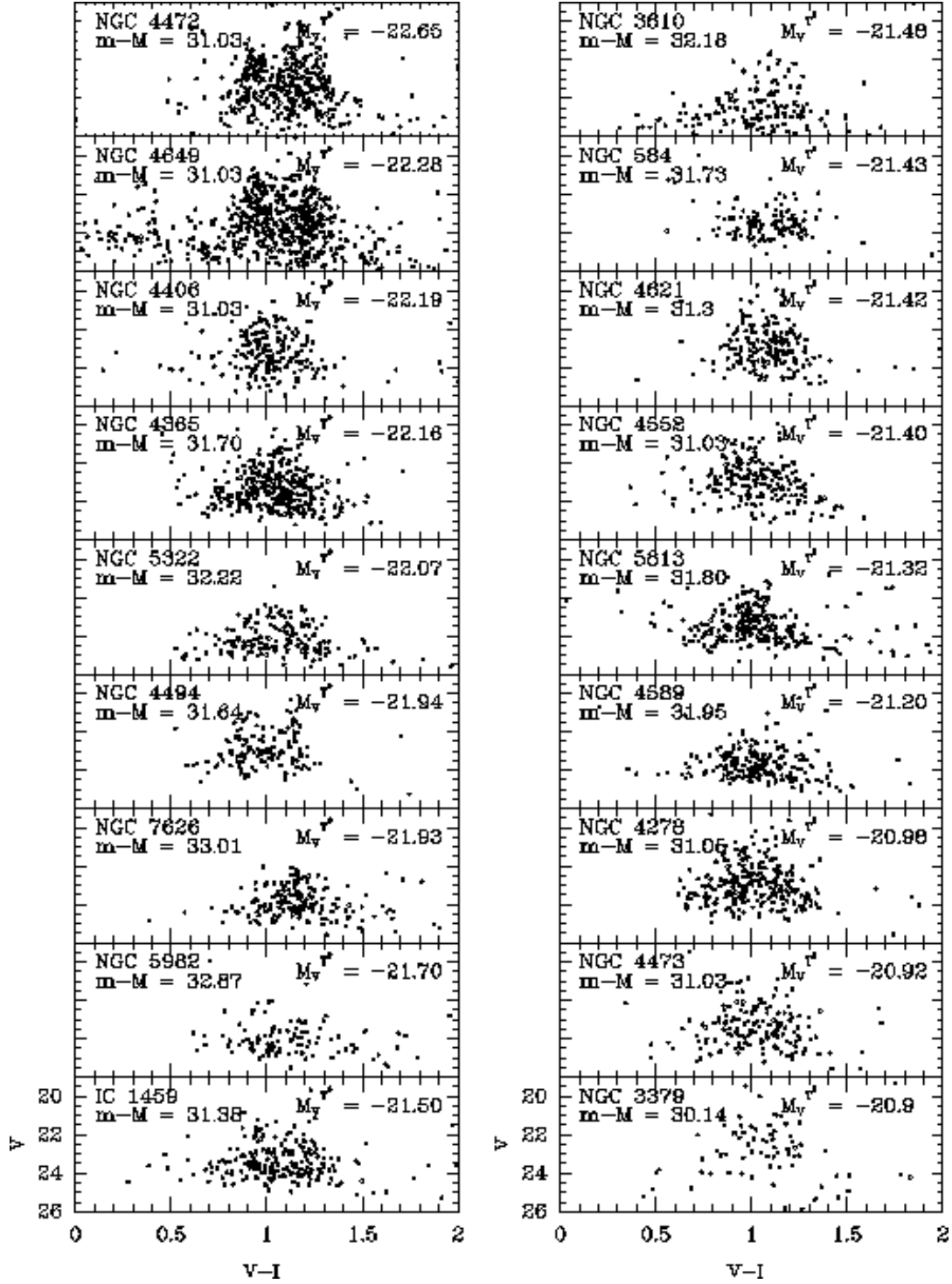}}
 \caption{The color-magnitude diagrams for the globular cluster candidates in the
program galaxies. Most of the candidates lie in a narrow range of color between
0.5$<$V-I$<$1.5. The distances  and  magnitudes are from Table 1 and the plots are sorted by the absolute magnitudes of the galaxy.  \label{fig1}}
\end{figure}
\clearpage
\centerline{\psfig{figure=fig1b.epsi,width=14.3cm}}
\clearpage

\begin{figure}
\centerline{\psfig{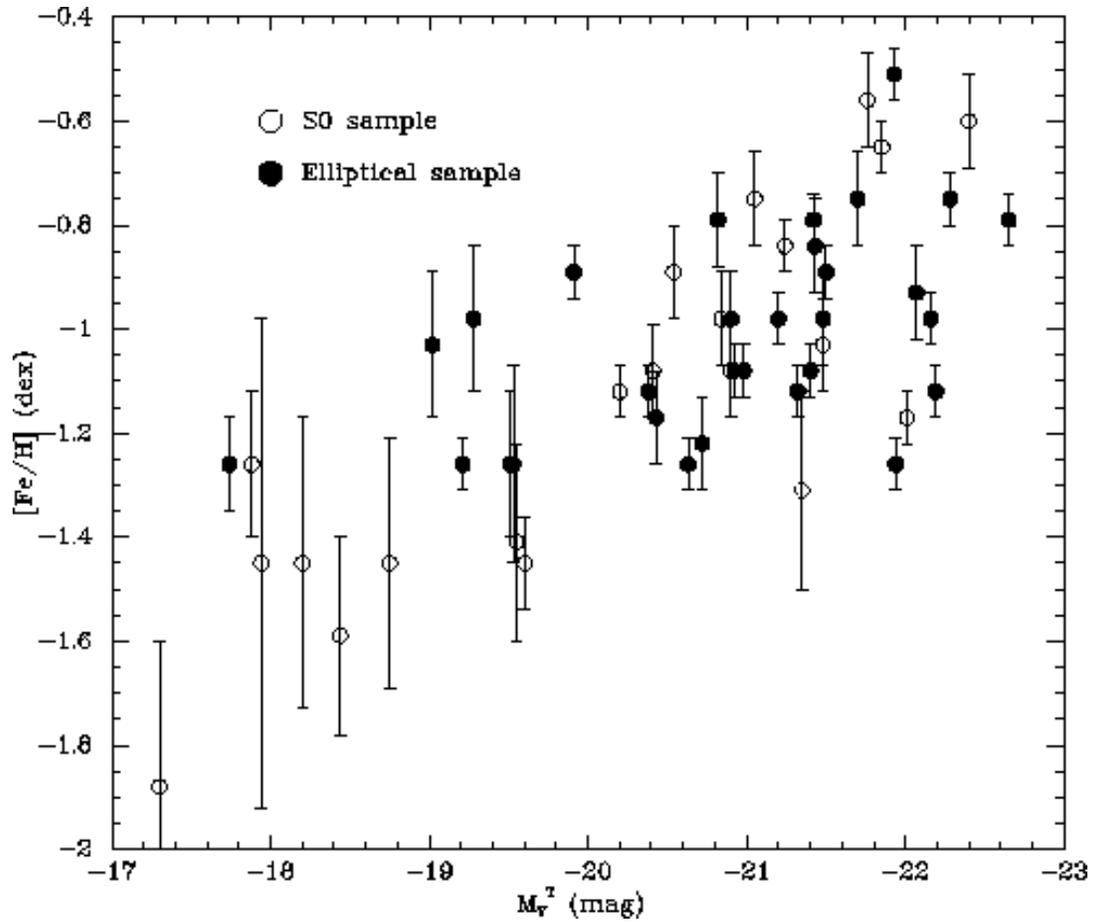}}
\caption{The average metallicity of the cluster systems  vs the absolute magnitude of the host galaxy.  \label{fig2}}
\end{figure}

\begin{figure}
\centerline{\psfig{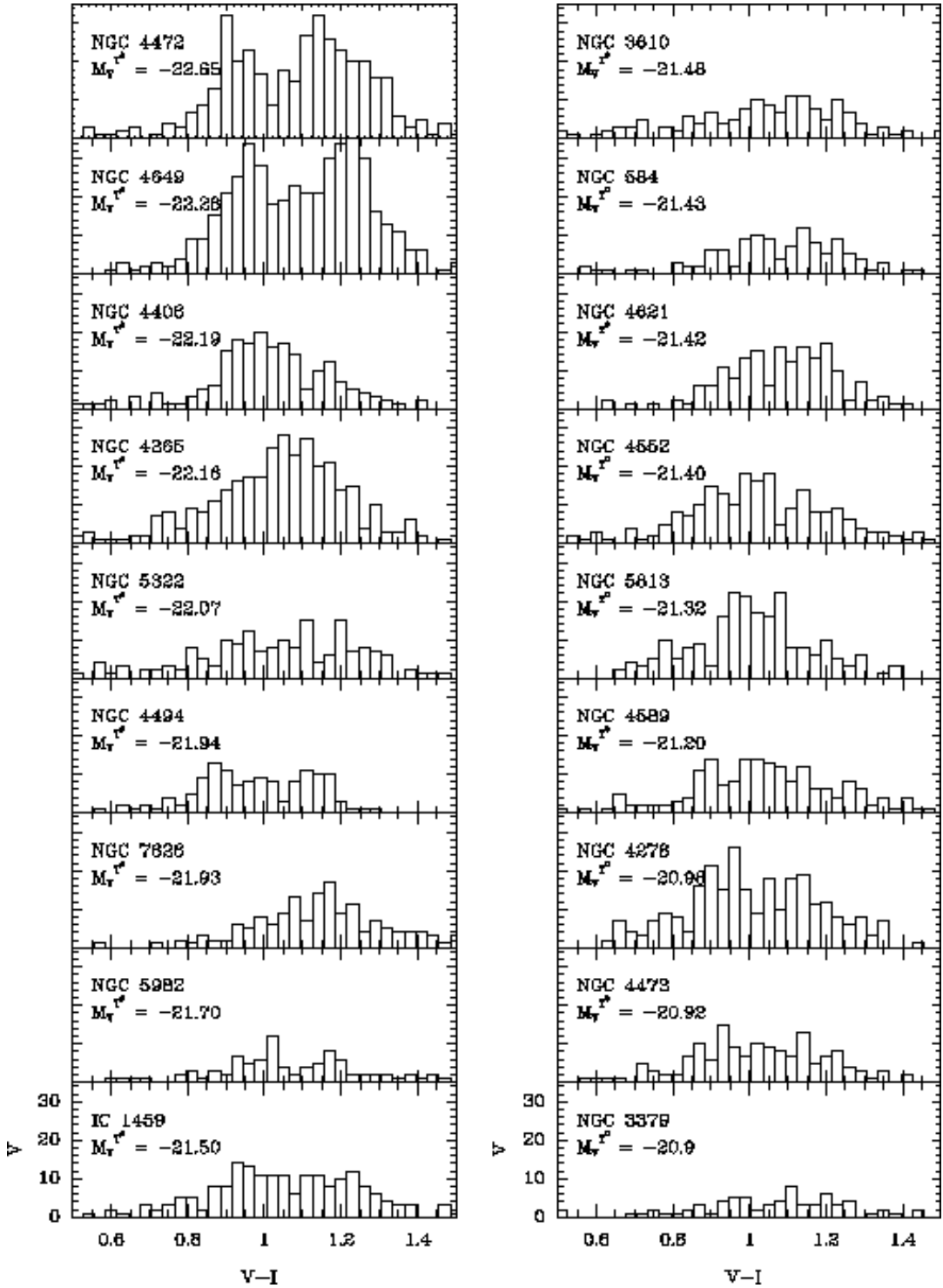}}
\caption{Color distributions of the globular cluster candidates. Clear evidence bimodality can be seen in a few galaxies. Statistical tests show evidence of bimodality in 17 of the 29 galaxies. \label{fig3}}
\end{figure}
\clearpage
\centerline{\psfig{figure=fig3b.epsi,width=14.3cm}}
\clearpage

\begin{figure}
\centerline{\psfig{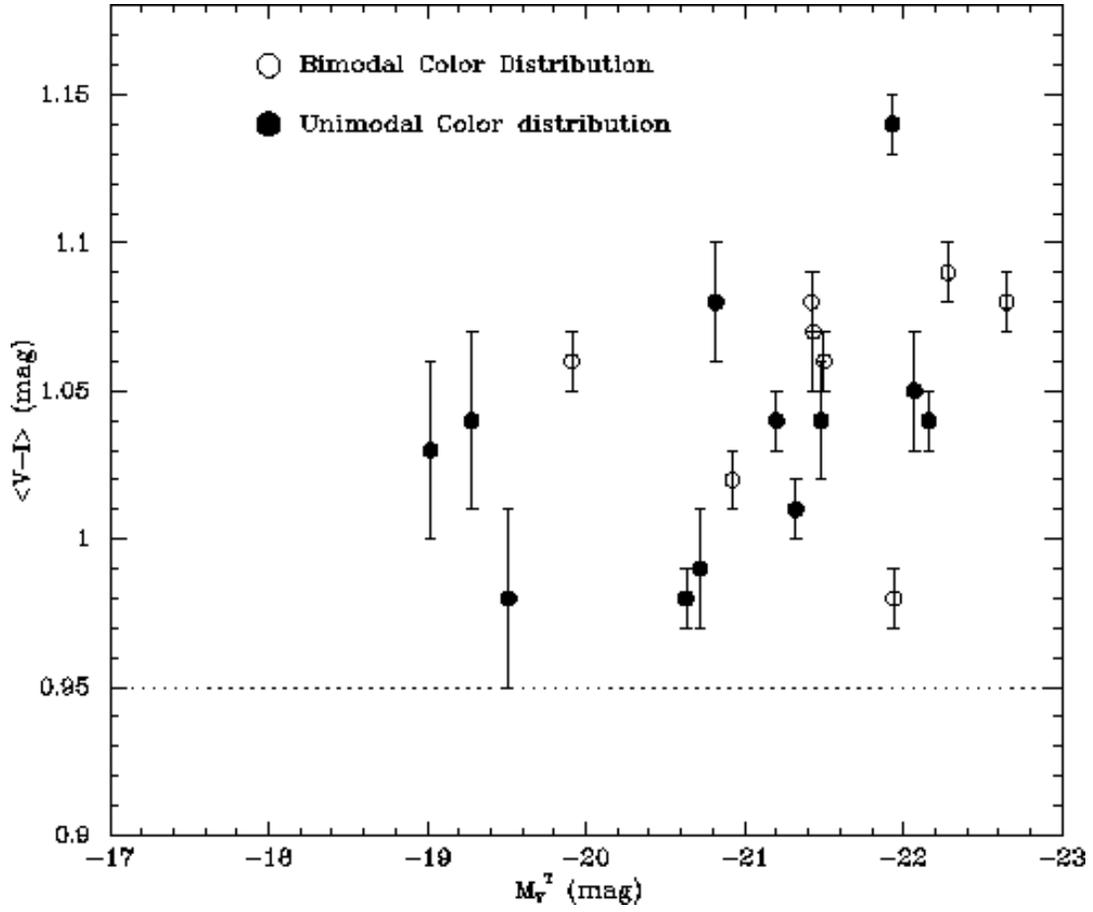}}
\caption{The mean color of the globular clusters with confirmed bimodality or confirmed lack of bimodality as a function of luminosity. No significant  differences are observed between the two populations. \label{fig4}}
\end{figure}

\begin{figure}
\centerline{\psfig{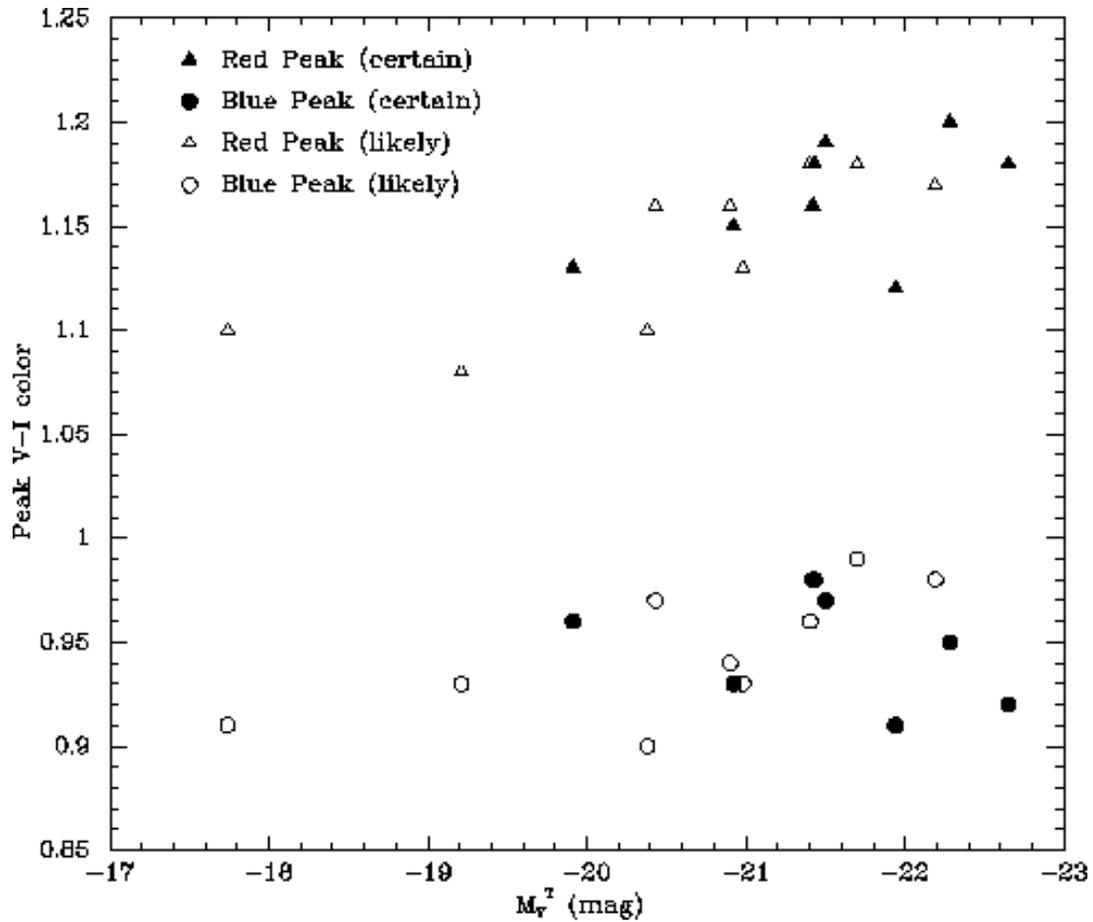}}
\caption{The peak color of the red and blue distributions vs absolute magnitude of the host (mass). \label{fig5}}
\end{figure}

\begin{figure}
\centerline{\psfig{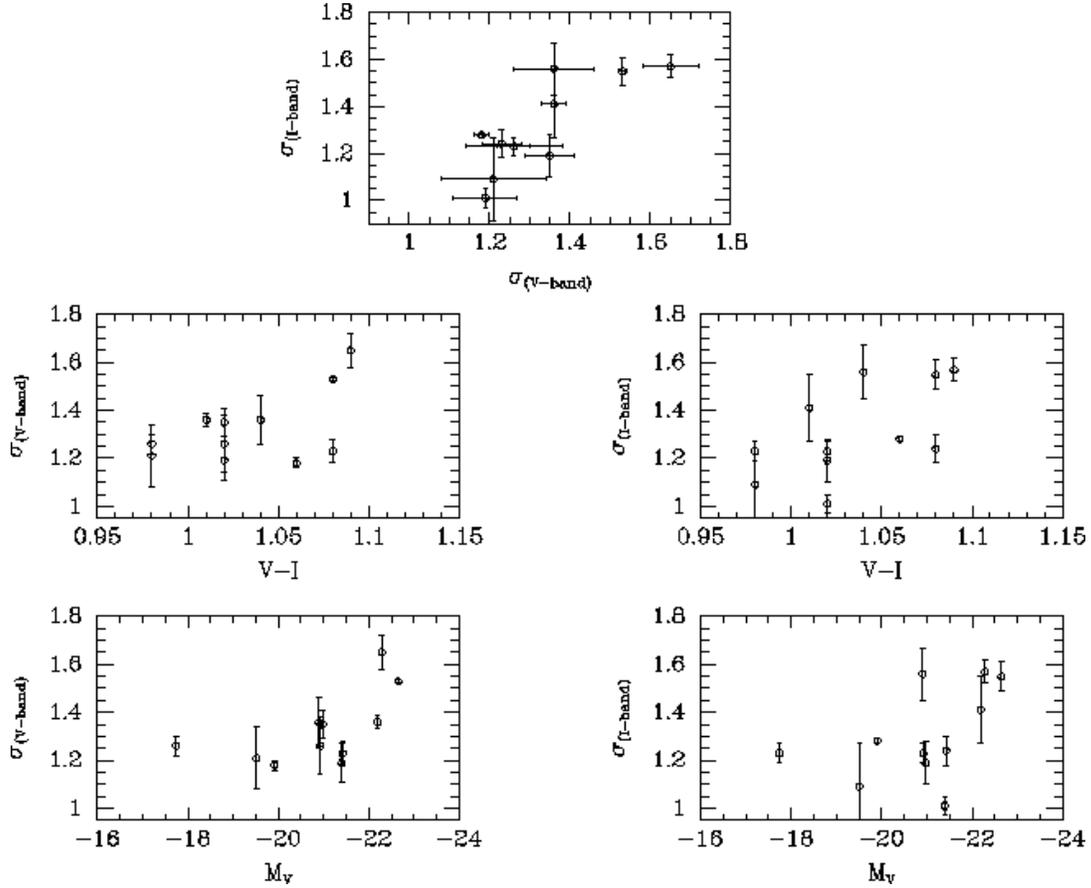}}
\caption{ Top panel: $\sigma_V$ vs $\sigma_I$ from the Gaussian fit to the
V and I-band GCLFs. Middle panels: Variation of $\sigma_V$ and $\sigma_I$ with 
the mean V-I color of the cluster system. Bottom panels: Variation of $\sigma_V$
 and $\sigma_I$ with the absolute magnitude (mass) of the host galaxy. The width
of the GCLF in the V and I bands appear to well correlated while there is a weak
trend of the more metal-rich clusters in more luminous galaxies having
wider GCLFs. 
\label{fig6}}
\end{figure}

\begin{figure}
\centerline{\psfig{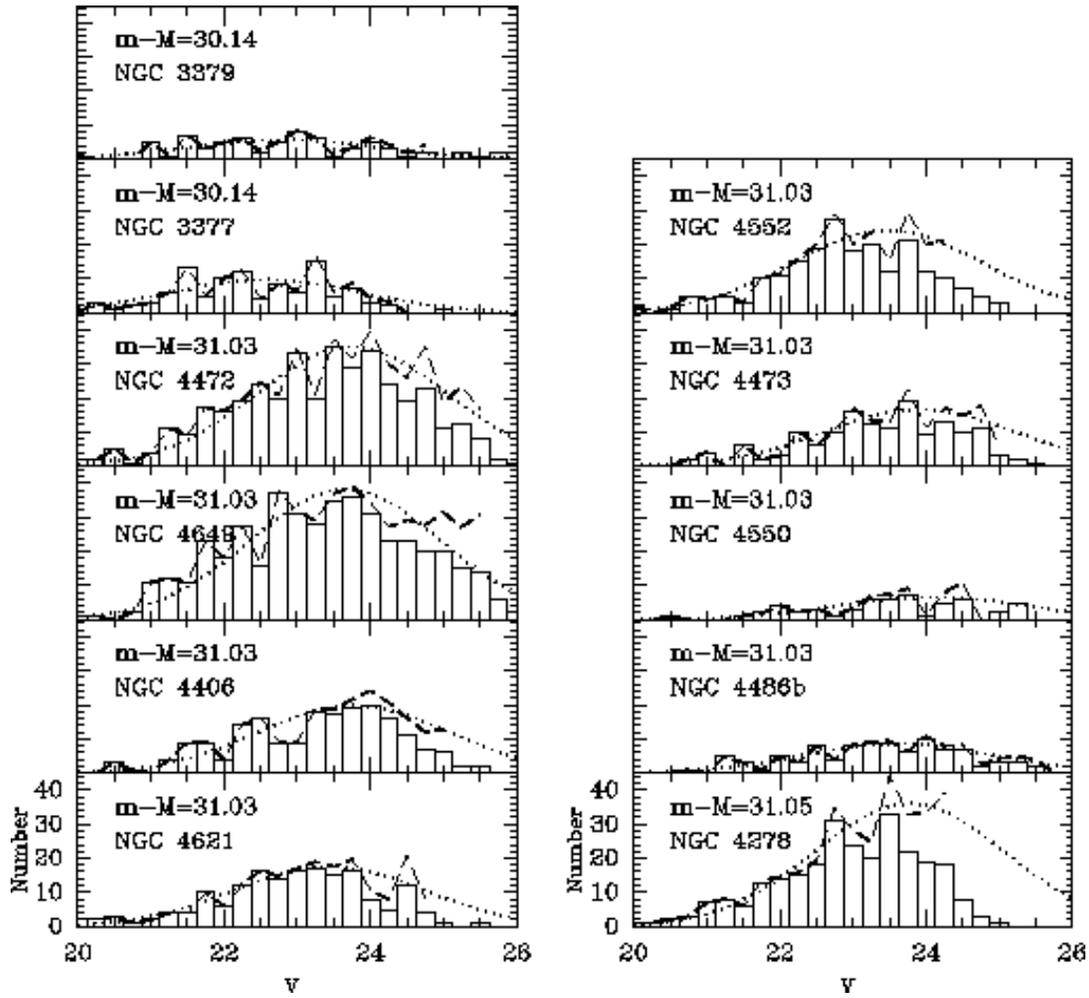}}
\caption{ a: The globular cluster luminosity function for the  systems in which the 50$\%$ completeness limit is fainter than the turnover luminosity. The dashed lines trace the 
completeness corrected distribution up to the 50$\%$ completeness limit. The dotted lines trace the best fit Gaussian with $\sigma$=1.3 mags.
b: The corresponding GCLFs in the I-band. The luminosity of the clusters are typically $\sim$1 mag brighter in I. \label{fig7}}
\end{figure}
\clearpage
\centerline{\psfig{figure=fig7b.epsi,width=14.3cm}}
\clearpage

\begin{figure}
\centerline{\psfig{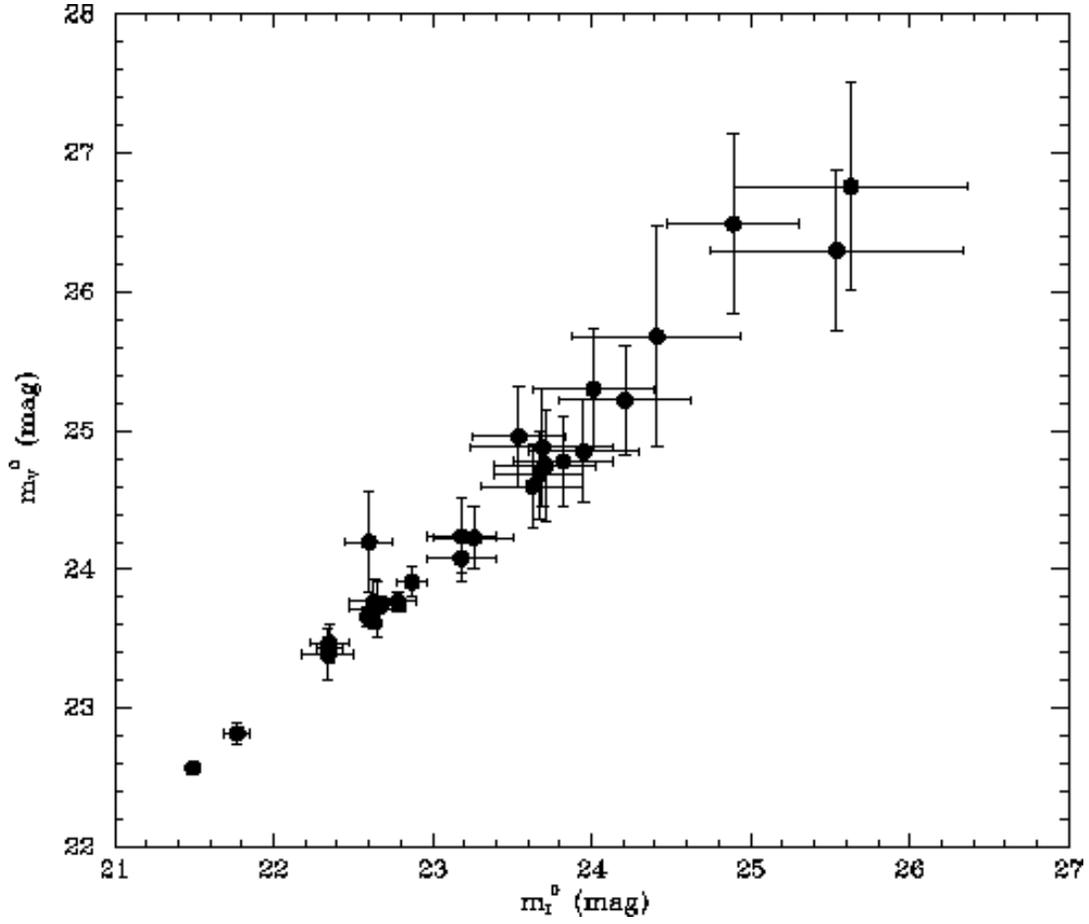}}
\caption{The turnover luminosity of the best fitting Gaussian to the GCLF in
 the V-band, m$_V^0$ vs the turnover luminosity in the I-band, m$_I^0$. The turnover luminosities in the two filters are very tightly correlated.   \label{fig8} }
\end{figure}

\begin{figure}
\centerline{\psfig{figure=fig9.epsi,width=10cm}}
\caption{ The difference between the turnovers in the V and I-bands vs the mean V-I color of the cluster system for variable $\sigma$ Gaussian fits (top) and  $\sigma$=1.3 Gaussian fits (bottom). A data point for M87 (Kundu et al. 1999)
has also been included in the plot.  m$_V^0$-m$_I^0$ increases with color
 (metallicity), as predicted by the Ashman, Conti, \& Zepf (1995). The dashed lines plot the best fit line through the data points.
\label{fig9}}
\end{figure}

\begin{figure}
\centerline{\psfig{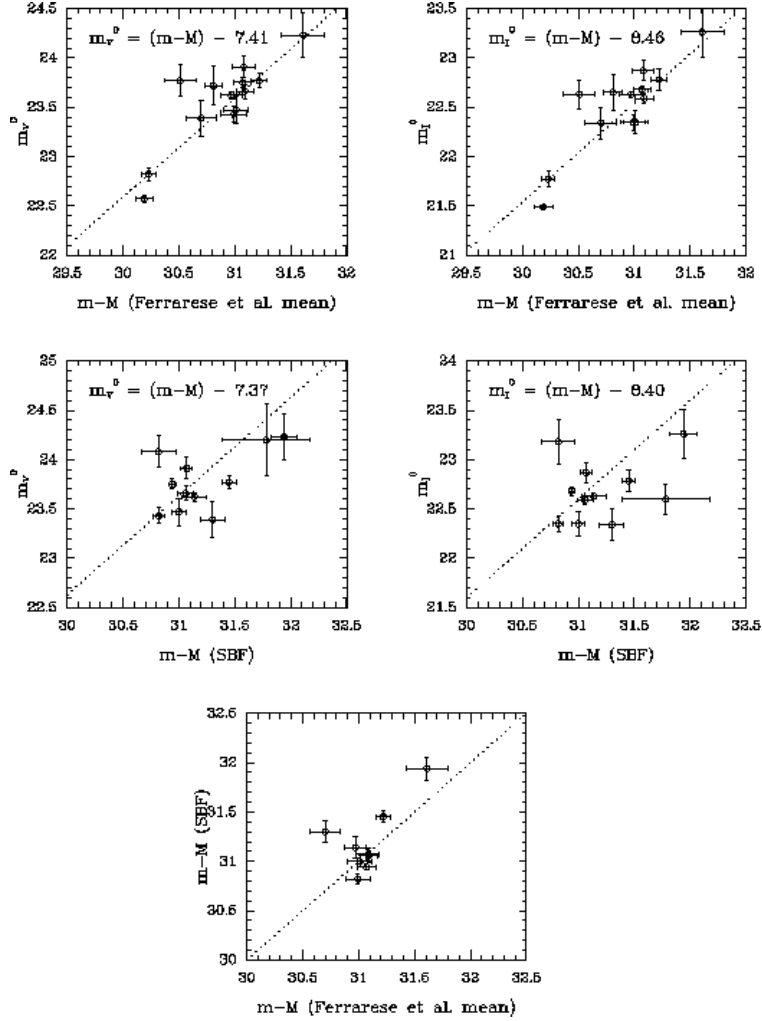}}
\caption{ Top panels: GCLF turnovers in the V and I-bands (for $\sigma$=1.3
 Gaussian fits)  compared with the distance
modulus from Ferrarese et al. (2000). The dotted lines trace the constant offset between the turnover and distance modulus reported in Table 5. Middle panels: Corresponding comparisons with the Neilsen et al. (1999) SBF distance measurements. Bottom panel: Comparison of the SBF measurements (Neilsen et al.)  to the Ferrarese et al. mean distance modulus. The GCLF turnover appears to be an excellent distance indicator with an
accuracy comparable with the SBF method.  
  \label{fig10} }
\end{figure}

\begin{figure}
\centerline{\psfig{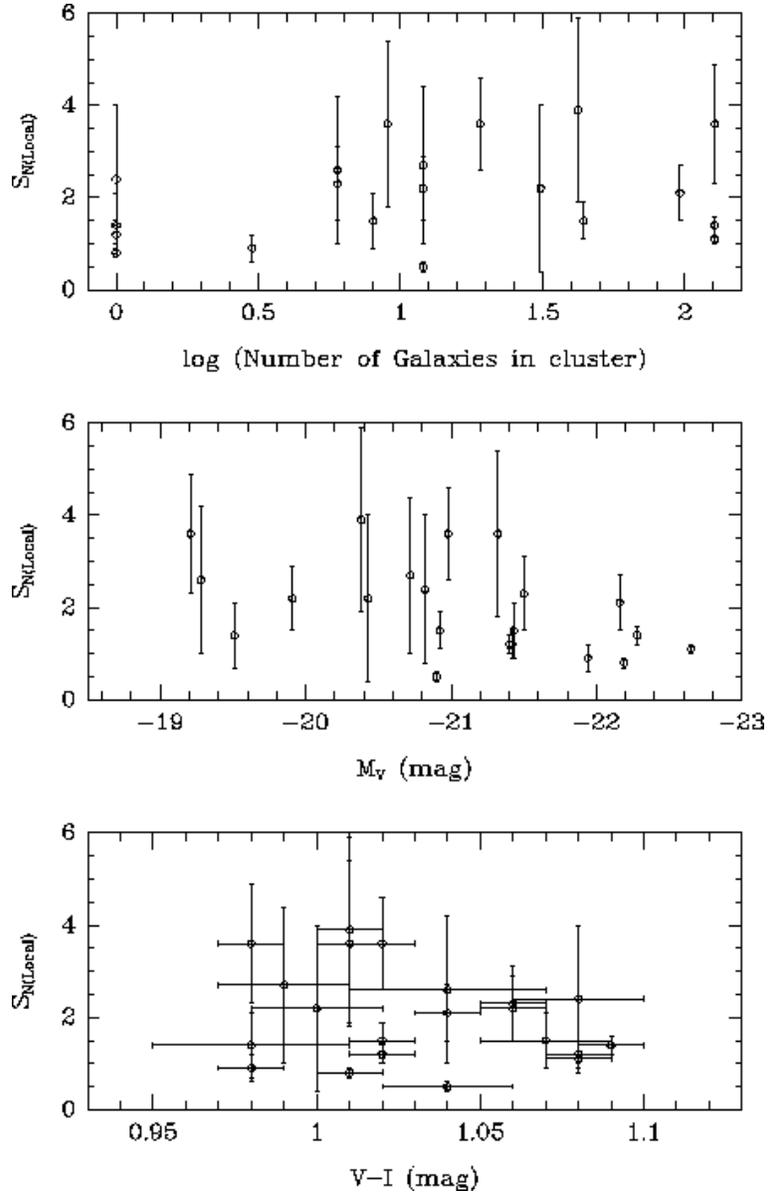}}
\caption{ Local specific frequency as a function of the log of the number of
galaxies in the host (galaxy) cluster (top), host galaxy luminosity (middle),
and mean clusters color (bottom).  \label{fig11} }
\end{figure}

\begin{figure}
\centerline{\psfig{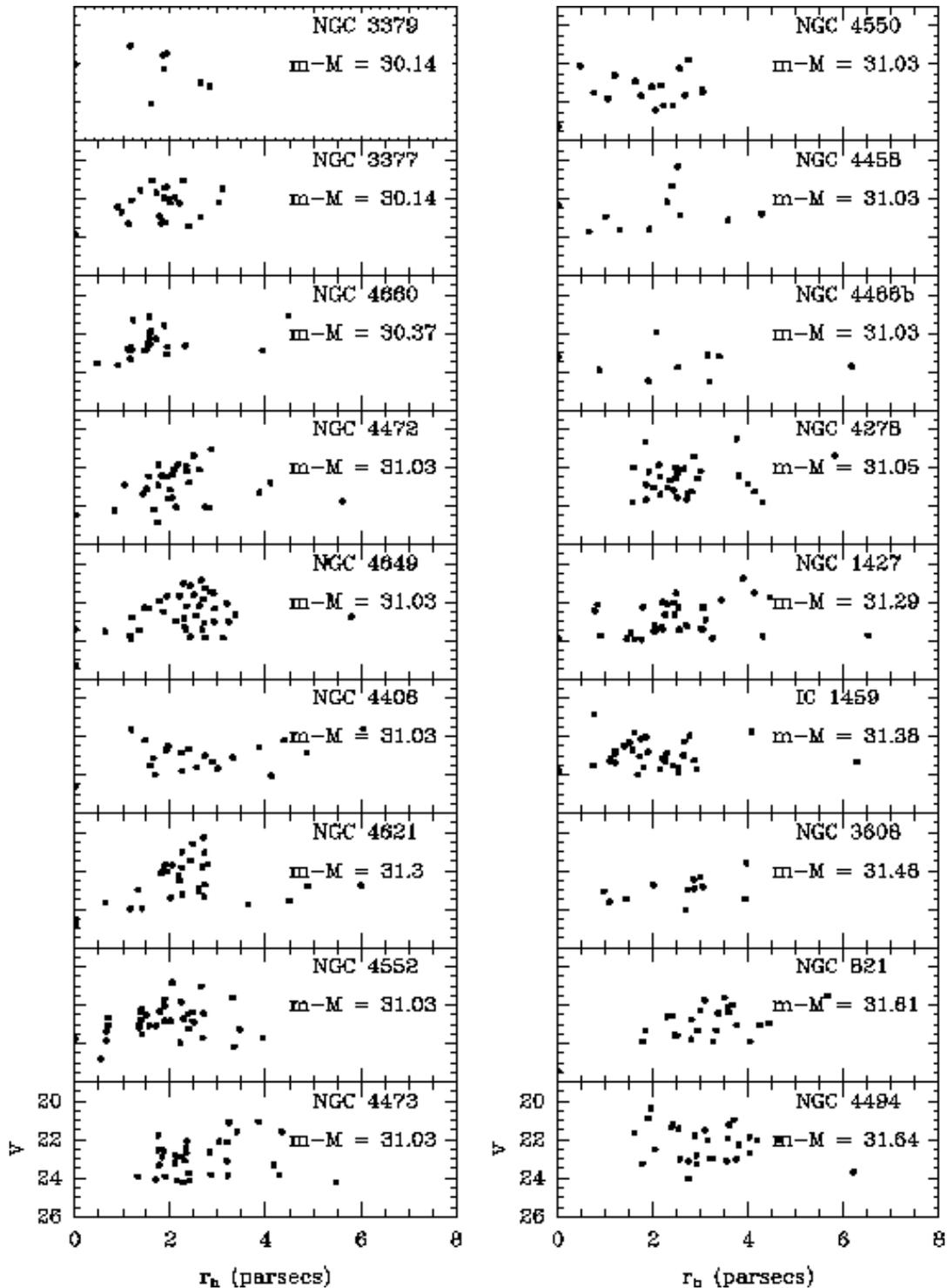}}
\caption{The half light radius  vs V-band magnitude of globular clusters in the PC. 
The half light radius of clusters in all the galaxies is less than 6 pc with a mean r$_h$$\approx$2.4 pc. There appears to be no magnitude dependence of the size of the clusters. The plots are sorted in order of distance to the host (and host luminosity in cases of degeneracy). \label{fig12}}
\end{figure}
\clearpage
\centerline{\psfig{figure=fig12b.epsi,width=14.3cm}}
\clearpage

\begin{figure}
\centerline{\psfig{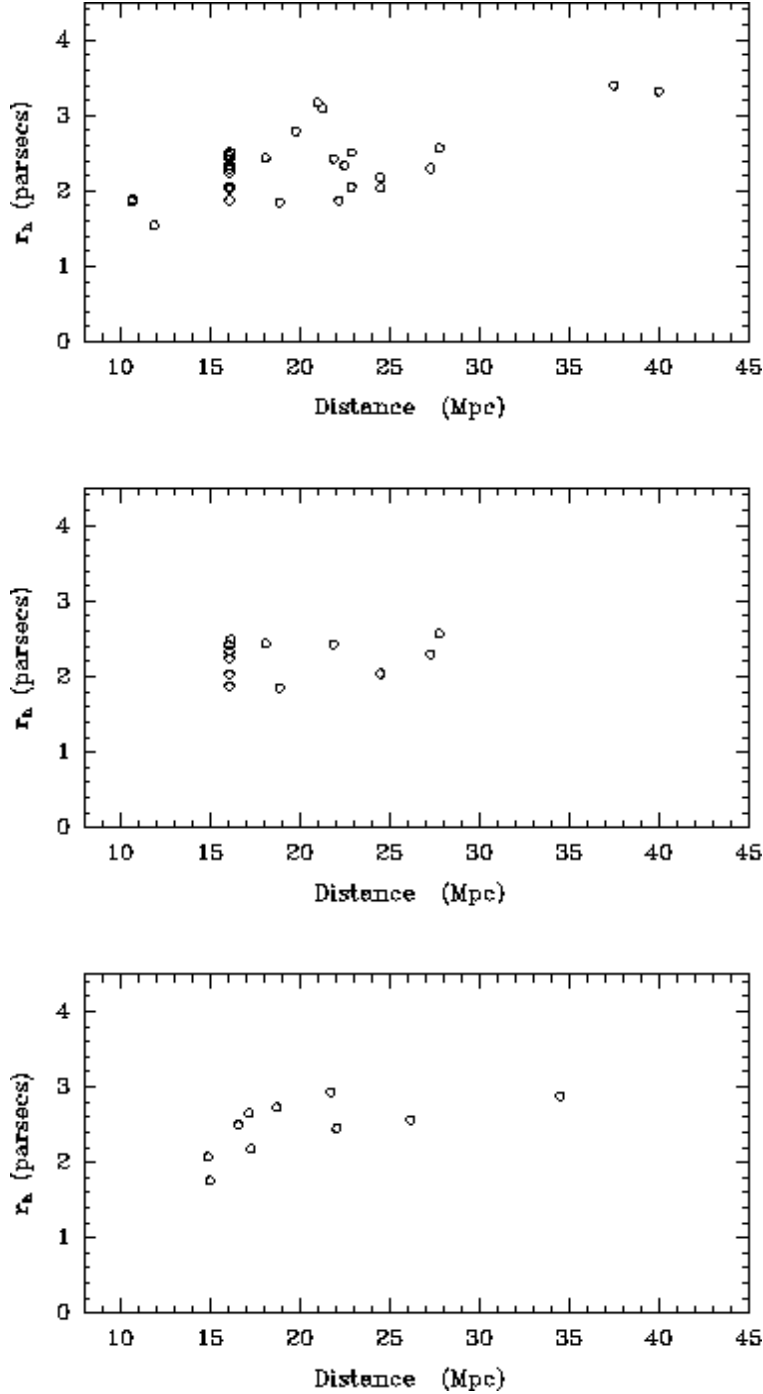}}
\caption{ Top: The median half light radius as a function of distance. The half-light radius appears to be nearly constant in all galaxies. The small distance 
dependent slope is most likely because of small error in the adopted PSF model. 
Middle: The median half light radius as a function of distance for galaxies with at least 30 cluster candidates in the PC. The scatter in the sizes is clearly much smaller than in the top panel. 
Bottom: The sizes and distances of the same set of candidates as in the middle panel using GCLF distances. The small distance dependent gradient is likely to be due to an error in the adopted PSF model. The median sizes are likely to be even more consistent in the absence of this effect. 
\label{fig13} }
\end{figure}

\begin{deluxetable}{llclclcll}
\tablenum{1}
\tablecaption{General Properties of the Elliptical Sample}
\startdata
    \\
\tableline
\tableline
Galaxy    & Morph. Typ           & Gp/Clstr & Hel Vel & (m-M) \tablenotemark{a} & Ref \tablenotemark{b} & A$_V$ \tablenotemark{c} & M$_V^{T}$ \tablenotemark{d}    \\
	  &		  &         & Km s$^{-1}$ & mag &   &  mag  & 	mag	 \\
(1)	  & (2)	  	& (3)	 & (4)	& (5)  & (6)  & (7)   & (8) \\	  
\tableline		 
NGC 4472  &  E2/S0(2)	& Virgo	 & 868	& 31.03& 1,2,3&	 0.00 &		-22.65	 \\
NGC 4649  &  E2		& Virgo	 & 1413 & 31.03& 1,2  &	 0.03 &		-22.28	 \\
NGC 4406  &  E3/S0(3)	& Virgo 	 & -227	& 31.03&1,2,3 &  0.08 &         -22.19   \\
NGC 4365  &  E3		& Virgo W  & 1240 & 31.70&1,3   &  0.00 &         -22.16   \\
NGC 5322  & E3-4;LIN	& N5322 Gp  & 1915 & 32.22& 5    &  0.00 &         -22.07   \\
NGC 4494  &  E1-2	& Coma II   & 1324 & 31.64& 5    &  0.04 &         -21.94   \\
NGC 7626  & E(pec)	& Pegasus   & 3423 & 33.01& 5    &  0.12 &         -21.93   \\
NGC 5982  & E3		& GH 158    & 2904 & 32.87& 5    &  0.01 &         -21.70   \\
IC 1459   &  E3		& I1459 Gp & 1691 & 31.38& 5    &  0.00 &         -21.50   \\
NGC 3610  &  E5		& N3642 Gp & 1787 & 32.18& 5    &  0.00 &         -21.48   \\
NGC 584	  &  E4		&  N584 Gp & 1875 & 31.73& 5    &  0.08 &         -21.43   \\
NGC 4621  &  E5		& Virgo 	 & 424  & 31.03& 1,2  &  0.05 &         -21.42   \\
NGC 4552  &  E;LIN      & Virgo 	 & 321  & 31.03&1,2,3 &  0.11 &         -21.40   \\
NGC 5813  & E1-2	& N5846 Gp  & 1924 & 31.80& 5    &  0.12 &         -21.32   \\
NGC 4589  &  E2		& N4291 Gp  & 1980 & 31.95& 5    &  0.02 &         -21.20   \\
NGC 4278  &E1-2;LIN    & Coma I   & 649  & 31.05& 5    &  0.08 &         -20.98   \\
NGC 4473  &  E5         & Virgo	 & 2240	& 31.03& 1,2  &  0.05 &         -20.92   \\
NGC 3379  &  E1		& Leo I    & 920 & 30.14 & 1,2  &  0.04 &         -20.90   \\
NGC 821	  &  E6?	&          & 1718 & 31.61& 5    &  0.12 &         -20.82   \\
NGC 3608  &  E2		& N3607 Gp & 1108 & 31.48& 5    &  0.00 &         -20.72   \\
NGC 4291  &  E		& N4291 Gp & 1757 & 31.95& 5    &  0.05 &         -20.63   \\
NGC 1439  &  E1		& Eridanus & 1670 & 31.76& 3,5  &  0.05 &         -20.43   \\
NGC 1427  &  E5		& Fornax   & 1416 & 31.29& 5    &  0.00 &         -20.38   \\
NGC 3377  &  E5-6	& Leo I    & 692 & 30.14 & 1,2  &  0.05 &         -19.91   \\
NGC 4550  &  SB0$^0$	& Virgo     & 381  & 31.03& 1    &  0.12 &         -19.51   \\
NGC 5845  &  E		& N5846 Gp? & 1450 & 31.80& 5    &  0.11 &         -19.28   \\
NGC 4660  &  E		& Virgo?    & 1097 & 30.37& 2    &  0.00 &         -19.21   \\
NGC 4458  &  E0-1	& Virgo 	 & 668 	& 31.03& 1,2  &  0.05 &         -19.02   \\
NGC 4486B &  cE0	& Virgo	 & 1486	& 31.03& 1    &  0.07 &         -17.74   \\
\enddata
\tablenotetext{} {(1) Galaxy : (2) Morphological classification : 
(3) Group/Cluster membership : (4) Heliocentric radial velocity : (5) Distance modulus : 
(6) References for distance modulus : (7) Galactic reddening in V-band : (8) Absolute magnitude
in the V-band }
\tablenotetext{a} { Velocity distances from Prugniel \& Simien (1996) }
\tablenotetext{b} { 1: Tonry, Blakeslee, Ajhar \& Dressler (1997) ; 2: Ajhar et al. (1997);
3: Jensen, Tonry \& Luppino (1998) ; 4: Neilsen, Tsvetanov \& Ford (1997); 
5: Prugniel \& Simien (1996) }
\tablenotetext{c} { Using A$_B$ from the NED extragalactic database and A$_V$=0.78A$_B$}
\tablenotetext{d} { Using m$_V^0$ from the NED extragalactic database and the distance modulus 
from column 5 }
    \\
\end{deluxetable}

\begin{deluxetable}{llcclclccl}
\tablenum{2}
\tablecaption{ Observing Log}
\startdata
\\
\tableline
\tableline
Galaxy    & Date	& t$_{exp}$(F555W) & t$_{exp}$(F814W) & Gain	  \\
	  &		& secs		& secs		& e$^-$ per ADU \\
(1)	  &	(2)	& (3)		& (4)		& (5) \\	
\tableline
NGC 4472  &  Feb 1995   & 1800		& 1800  & 7  \\
NGC 4649  &  Apr 1996	& 2100 		& 2500  & 7  \\
NGC 4406  &  Nov 1994	& 1500		& 1500  & 7  \\
NGC 4365  &  May 1994	& 1000		& 460   & 15 \\
NGC 5322  &  Nov 1994	& 1000		& 460   & 7  \\
NGC 4494  &  May 1994	& 1000		& 460   & 15 \\
NGC 7626  &  May 1994	& 1000		& 460   & 15 \\
NGC 5982  &  May 1994	& 1000		& 460   & 15 \\
IC 1459	  &  Jun 1994	& 1000		& 460   & 15 \\
NGC 3610  &  Dec 1994   & 1200		& 1000  & 7  \\
NGC 584	  &  Jan 1996   & 1400		& 1380	& 15 \\
NGC 4621  &  Feb 1995	& 1050		& 1050  & 7  \\
NGC 4552  &  Nov 1995 	& 1000		& 460   & 7  \\
NGC 5813  &  May 1994	& 1000		& 460   & 15 \\
NGC 4589  &  May 1994 	& 1000		& 460   & 15 \\
NGC 4278  &  May 1994   & 1000		& 460   & 15 \\
NGC 4473  &  May 1996	& 1800		& 2000  & 7  \\
NGC 3379  &  Nov 1994	& 1500		& 1200  & 7  \\
NGC 821	  &  Jan 1996   & 1400		& 1380 	& 15 \\
NGC 3608  &  May 1994   & 1000		& 460   & 15 \\
NGC 4291  &  Nov 1995   & 1400		& 1380  & 15 \\
NGC 1439  &  Jul 1994   & 1000		& 460   & 7  \\
NGC 1427  &  May 1994	& 1000		& 460   & 15 \\
NGC 3377  &  Apr 1995	& 1050		& 1050  & 7  \\
NGC 4550  &  Dec 1994	& 1200		& 1200  & 7  \\
NGC 5845  &  Jan 1996	& 1200		& 1040  & 15 \\
NGC 4660  &  Dec 1994	& 920		& 800   & 15 \\
NGC 4458  &  Feb 1995	& 1200		& 1040  & 15 \\
NGC 4486B &  Nov 1995	& 1800		& 2000  & 7  \\

\enddata
\tablenotetext{} {(1) Galaxy : (2) Observation date : (3) Exposure time in filter F555W : (4) Exposure time in filter F814W :
(5) CCD analog-to-digital gain in electrons per ADU }
\\
\end{deluxetable}

\begin{deluxetable}{llllll}
\tablenum{3}
\tablecaption{ Elliptical Sample: Color and Metallicity Distributions}
\startdata
\\
\tableline
\tableline
Galaxy        & No$_{cand}$ & $<$V-I$>$     &  [Fe/H]	& Bimodal? & V-I peak  \\
	  &		    &	mag	  &  dex        &  & mag  \\
(1)	  & (2)	  	& (3)	 & (4)	& (5)   & (6) \\	  
\tableline		
NGC 4472 & 381 	&  1.08 (0.01) &  -0.79 (0.05) & Y &  0.92,1.18  \\
NGC 4649 & 445 	&  1.09 (0.01) &  -0.75 (0.05) & Y &  0.95,1.2  \\
NGC 4406 & 196 	&  1.01 (0.01) &  -1.12 (0.05) & L &  0.98,1.17  \\
NGC 4365 & 325 	&  1.04 (0.01) &  -0.98 (0.05) & N &      \\
NGC 5322 & 171 	&  1.05 (0.02) &  -0.93 (0.09) & N &      \\
NGC 4494 & 131 	&  0.98 (0.01) &  -1.26 (0.05) & Y &  0.91,1.12  \\
NGC 7626 & 152 	&  1.14 (0.01) &  -0.51 (0.05) & N &      \\
NGC 5982 &  89 	&  1.09 (0.02) &  -0.75 (0.09) & L &  0.99,1.18  \\
IC 1459  & 186 	&  1.06 (0.01) &  -0.89 (0.05) & Y &  0.97,1.19  \\
NGC 3610 & 147 	&  1.04 (0.02) &  -0.98 (0.09) & N &      \\
NGC  584 & 115 	&  1.07 (0.02) &  -0.84 (0.09) & L &  0.98,1.18  \\
NGC 4621 & 172 	&  1.08 (0.01) &  -0.79 (0.05) & Y &  0.98,1.16  \\
NGC 4552 & 210 	&  1.02 (0.01) &  -1.08 (0.05) & L &  0.96,1.18  \\
NGC 5813 & 213 	&  1.01 (0.01) &  -1.12 (0.05) & N &      \\
NGC 4589 & 179 	&  1.04 (0.01) &  -0.98 (0.05) & N &      \\
NGC 4278 & 267 	&  1.02 (0.01) &  -1.08 (0.05) & L &  0.93,1.13  \\
NGC 4473 & 150 	&  1.02 (0.01) &  -1.08 (0.05) & Y &  0.93,1.15  \\
NGC 3379 &  70	&  1.04 (0.02) &  -0.98 (0.09) & L &  0.94,1.16  \\
NGC  821 & 105 	&  1.08 (0.02) &  -0.79 (0.09) & N &      \\
NGC 3608 & 103 	&  0.99 (0.02) &  -1.22 (0.09) & N &      \\
NGC 4291 & 143 	&  0.98 (0.01) &  -1.26 (0.05) & N &      \\
NGC 1439 &  95	&  1.00 (0.02) &  -1.17 (0.09) & L &  0.97,1.16  \\
NGC 1427 & 163 	&  1.01 (0.01) &  -1.12 (0.05) & L &   0.9,1.1  \\
NGC 3377 & 106 	&  1.06 (0.01) &  -0.89 (0.05) & L &  0.96,1.13  \\
NGC 4550 &  55	&  0.98 (0.03) &  -1.26 (0.14) & N &      \\
NGC 5845 &  41	&  1.04 (0.03) &  -0.98 (0.14) & N &      \\
NGC 4660 & 101 	&  0.98 (0.01) &  -1.26 (0.05) & L &  0.93,1.08  \\
NGC 4458 &  43	&  1.03 (0.03) &  -1.03 (0.14) & N &      \\
NGC 4486B & 97	&   0.98 (0.02) &  -1.26 (0.09) & L &  0.91,1.1  \\
\tableline
Avg$^\dag$	&    & 1.04$\pm$0.04 (0.01)	&  -1.0$\pm$0.19 (0.04) & & \\
\enddata
\tablenotetext{} {(1) Galaxy (2) Number of globular cluster candidates (3) Mean V-I color 
(4) Mean metallicity using eq. 2.1 (5) Bimodal? Y=certain, 
L=very likely, N=no (6) V-I color of the blue and red peak 
}
\tablenotetext{\dag} {Excluding NGC 4550 which is a S0 galaxy.}
\end{deluxetable}

\begin{deluxetable}{llllllllll}
\tablenum{4}
\tablecaption{ Elliptical Sample: Turnover Magnitudes}
\startdata
    \\
\tableline
\tableline
	  & 	  & \multicolumn{2}{c}{$\sigma_{V}$ variable} & $\sigma_{V}$=1.3 & & \multicolumn{2}{c}{$\sigma_{I}$ variable} & $\sigma_{I}$=1.3 \\
\cline{3-4} \cline{7-8}\\
Galaxy    &  m-M  &  m$_V^0$ 	   & $\sigma_{V}$  &	m$_V^0$	     & V$_{compl}$ & m$_I^0$   &  $\sigma_{I}$ &  m$_I^0$ 	& I$_{compl}$          \\
	  &	mag	  &  mag        & mag  & mag & mag \\
(1)	  & (2)	  & (3)	 	   & (4)	   & (5) 	     & (6)  & 	(7)	       &	(8)    &	(9)	& (10) \\	  
\tableline		
NGC 3379 &  30.14 & 22.84$\pm$0.1 & 1.36$\pm$0.1 &  22.82$\pm$0.07 & 24.7 &   21.92$\pm$0.12 & 1.56$\pm$0.11 & 21.77$\pm$0.08 & 23.6  \\
NGC 3377 &  30.14 & 22.55$\pm$0.02 & 1.18$\pm$0.02 &  22.57$\pm$0.03 & 24.7 &   21.49$\pm$0.02 & 1.28$\pm$0.01 & 21.49$\pm$0.02 & 23.7  \\
NGC 4660 &  30.37 &                &               &  23.39$\pm$0.18 & 23.9 &                  &               & 22.34$\pm$0.16 & 22.9  \\
NGC 4472 &  31.03 & 23.84$\pm$0.03 & 1.53$\pm$0.01 &  23.75$\pm$0.05 & 25.3 &   22.78$\pm$0.06 & 1.55$\pm$0.06 & 22.68$\pm$0.04 & 24.3  \\
NGC 4649 &  31.03 &  23.8$\pm$0.08 & 1.65$\pm$0.07 &  23.66$\pm$0.07 & 25.4 &   22.61$\pm$0.05 & 1.57$\pm$0.05 & 22.59$\pm$0.05 & 24.3  \\
NGC 4406 &  31.03 & 23.79$\pm$0.03 & 1.36$\pm$0.03 &  23.77$\pm$0.07 & 24.9 &   22.86$\pm$0.21 & 1.41$\pm$0.14 & 22.78$\pm$0.11 & 23.8  \\
NGC 4621 &  31.03 & 23.39$\pm$0.05 & 1.23$\pm$0.05 &  23.43$\pm$0.08 & 24.8 &   22.31$\pm$0.06 & 1.24$\pm$0.06 & 22.35$\pm$0.08 & 23.8  \\
NGC 4552 &  31.03 & 23.37$\pm$0.11 & 1.19$\pm$0.08 &  23.47$\pm$0.14 & 24.2 &   22.17$\pm$0.03 & 1.01$\pm$0.04 & 22.35$\pm$0.12 & 23.3  \\
NGC 4473 &  31.03 &  23.9$\pm$0.15 & 1.26$\pm$0.12 &  23.91$\pm$0.11 & 25.1 &   22.84$\pm$0.09 & 1.23$\pm$0.04 & 22.87$\pm$0.1 &  24.  \\
NGC 4550 &  31.03 & 24.01$\pm$0.17 & 1.21$\pm$0.13 &  24.08$\pm$0.16 & 24.6 &   22.97$\pm$0.25 & 1.09$\pm$0.18 & 23.18$\pm$0.22 & 23.4  \\
NGC 4458 &  31.03 &                &               &   24.2$\pm$0.36 & 24.3 &                  &               &  22.6$\pm$0.15 & 23.4  \\
NGC 4486B&  31.03 & 23.61$\pm$0.04 & 1.26$\pm$0.04 &  23.62$\pm$0.04 & 25.5 &   22.61$\pm$0.02 & 1.23$\pm$0.04 & 22.63$\pm$0.03 & 24.5  \\
NGC 4278 &  31.05 & 23.81$\pm$0.09 & 1.35$\pm$0.06 &  23.77$\pm$0.16 & 24.1 &   22.54$\pm$0.12 & 1.19$\pm$0.09 & 22.63$\pm$0.15 & 23.2  \\
NGC 1427 &  31.29 &                &               &  24.24$\pm$0.27 & 24.2 &                  &               & 23.18$\pm$0.22 & 23.3  \\
IC 1459  &  31.38 &                &               &   24.6$\pm$0.3 & 24.3 &                  &               & 23.63$\pm$0.32 & 23.3  \\
NGC 3608 &  31.48 &                &               &  24.75$\pm$0.4 & 24.2 &                  &               & 23.71$\pm$0.32 & 23.3  \\
NGC  821 &  31.61 &                &               &  24.88$\pm$0.43 & 24.1 &                  &               & 23.69$\pm$0.45 & 23.1  \\
NGC 4494 &  31.64 &                &               &  23.72$\pm$0.2 & 24.1 &                  &               & 22.65$\pm$0.18 & 23.1  \\
NGC 4365 &   31.7 &                &               &  24.23$\pm$0.23 & 24.3 &                  &               & 23.26$\pm$0.25 & 23.3  \\
NGC  584 &  31.73 &                &               &  24.96$\pm$0.36 & 24.3 &                  &               & 23.54$\pm$0.29 & 23.2  \\
NGC 1439 &  31.76 &                &               &  24.68$\pm$0.31 & 24.8 &                  &               & 23.67$\pm$0.28 & 23.8  \\
NGC 5813 &   31.8 &                &               &  24.78$\pm$0.32 & 24.3 &                  &               & 23.82$\pm$0.31 & 23.3  \\
NGC 5845 &   31.8 &                &               &  24.85$\pm$0.37 & 24.4 &                  &               & 23.95$\pm$0.35 & 23.4  \\
NGC 4589 &  31.95 &                &               &  25.22$\pm$0.39 & 24.3 &                  &               & 24.21$\pm$0.41 & 23.3  \\
NGC 4291 &  31.95 &                &               &   25.3$\pm$0.44 & 24.1 &                  &               & 24.01$\pm$0.38 & 23.2  \\
NGC 3610 &  32.18 &                &               &  26.49$\pm$0.65 & 24.9 &                  &               & 24.89$\pm$0.41 & 23.9  \\
NGC 5322 &  32.22 &                &               &   26.3$\pm$0.58 & 24.6 &                  &               & 25.54$\pm$0.8 & 23.5  \\
NGC 5982 &  32.87 &                &               &  26.76$\pm$0.75 & 24.2 &                  &               & 25.63$\pm$0.73 & 23.2  \\
NGC 7626 &  33.01 &                &               &  25.68$\pm$0.8 & 24.1 &                  &               & 24.41$\pm$0.53 &  23.  \\
\tableline
Avg 	 &	  &		   & 1.32$\pm$0.15 & 		     &	    &	 & 1.30$\pm$0.19 \\
\enddata
\tablenotetext{} {(1) Galaxy : (2) Distance modulus from table 1 : 
(3) V-band turnover ($\sigma$ variable Gaussian fit) : (4) $\sigma$$_V$ : (5) V-band turnover ($\sigma$ = 1.3) : (6) V-band 50$\%$ completeness : (7) I-band
turnover ($\sigma$ variable Gaussian fit) : (8) $\sigma$$_I$ : (9) I-band turnover ($\sigma$ = 1.3) : (10) I-band 50$\%$ completeness }
\end{deluxetable}

\begin{deluxetable}{lllllll}
\tablenum{5}
\tablecaption{ GCLF Turnover }
\startdata
    \\
\tableline
\tableline
    & Ferrarese et al mean distance    &   SBF (Neilsen)   \\
\tableline		
  & \multicolumn{2}{c}{Unweighted mean GCLF turnover$^\dag$} \\
\cline{2-3}\\
\underline{ $\sigma$ = 1.3 GCLF fits}  \\
M$_V^0$ & -7.45$\pm$0.11 [7]	& -7.44$\pm$0.18 [5]	\\
M$_I^0$ & -8.50$\pm$0.14 [6]	& -8.43$\pm$0.11 [4]	\\
\underline{Variable $\sigma$ GCLF fits} 	\\
M$_V^0$ & -7.32$\pm$0.31 [7]	& -7.40$\pm$0.22 [5]	\\
M$_I^0$ & -8.51$\pm$0.23 [7]	& -8.45$\pm$0.24 [6]	\\
\\
\underline{SBF vs Ferrarese et al mean distance$^\dag$} & -0.02$\pm$0.14 [6]	\\
  & \multicolumn{2}{c}{ Weighted mean GCLF turnover} \\
\cline{2-3}\\
\underline{ $\sigma$ = 1.3 GCLF fits}  \\
M$_V^0$ & \underline{-7.41(0.03)$^\ddag$ [13]}	& -7.37(0.03) [11]	\\
M$_I^0$ & \underline{-8.46(0.03)$^\ddag$ [13]}	& -8.40(0.03) [11]	\\
\\
\tableline
\enddata
\tablenotetext{} {Note: Following the convention in the rest of the paper the
numbers quoted after the $\pm$ symbol are the standard deviations, while those
within parantheses are the uncertainties in the mean.  The integers within square brackets are the number of galaxies used to calculate the mean. See below for
details. }
\tablenotetext{\dag} { For the unweighted mean calculations, the GCLF values are limited to those with an uncertainty
 less than 0.1 mag. Also, for the comparison of the SBF vs Ferrarese et al mean distance only measurements with an uncertainty less than 0.1 mag have been considered.}
\tablenotetext{\ddag} { We adopt the weighted mean turnover with respect to the Ferrarese et al mean distance as the most reliable measurement of the absolute value of the GCLF peak. }
\end{deluxetable}

\begin{deluxetable}{lllllll}
\tablenum{6}
\tablecaption{ Comparison of GCLF distances with other methods }
\startdata
    \\
\tableline
\tableline
Galaxy$^\dag$    &  m-M    &   m-M$_{SBF}$      & m-M$_{Mean}$	  & m-M$_{V-GCLF}$  & m-M$_{I-GCLF}$ &  $<$m-M$>$$_{GCLF}$      \\
	  &	mag	  &  mag        & mag  & mag & mag \\
(1)	  & (2)	  	& (3)	 & (4)	& (5) & (6) & (7) \\	  
\tableline	

NGC  3379 &  30.14 &                 &   30.23$\pm$0.06 &    30.23$\pm$0.08 &   30.23$\pm$0.09 &  30.23$\pm$0.06\\
NGC  3377 &  30.14 &                 &   30.19$\pm$0.08 &    29.98$\pm$0.04 &   29.95$\pm$0.04 &  29.96$\pm$0.03\\
NGC  4660 &  30.37 &  31.3$\pm$0.11  &    30.7$\pm$0.14 &     30.8$\pm$0.18 &    30.8$\pm$0.16 &   30.8$\pm$0.12\\
NGC  4472 &  31.03 &  30.94$\pm$0.03 &   31.07$\pm$0.08 &    31.16$\pm$0.06 &   31.14$\pm$0.05 &  31.15$\pm$0.04\\
NGC  4649 &  31.03 &  31.06$\pm$0.07 &   31.09$\pm$0.08 &    31.07$\pm$0.08 &   31.05$\pm$0.06 &  31.06$\pm$0.05\\
NGC  4406 &  31.03 &  31.45$\pm$0.06 &   31.22$\pm$0.07 &    31.18$\pm$0.08 &   31.24$\pm$0.11 &  31.21$\pm$0.07\\
NGC  4621 &  31.03 &  30.82$\pm$0.05 &   30.99$\pm$0.11 &    30.84$\pm$0.09 &   30.81$\pm$0.09 &  30.82$\pm$0.06\\
NGC  4552 &  31.03 &  31.$\pm$0.06   &   31.01$\pm$0.11 &    30.88$\pm$0.14 &   30.81$\pm$0.12 &  30.84$\pm$0.09\\
NGC  4473 &  31.03 &  31.07$\pm$0.05 &   31.08$\pm$0.1  &    31.32$\pm$0.11 &   31.33$\pm$0.1 &  31.32$\pm$0.08\\
NGC  4550 &  31.03 &  30.82$\pm$0.15 &                  &    31.49$\pm$0.16 &   31.64$\pm$0.22 &  31.56$\pm$0.14\\
NGC  4458 &  31.03 &  31.78$\pm$0.39 &                  &    31.61$\pm$0.36 &   31.06$\pm$0.15 &  31.33$\pm$0.2\\
NGC  4486B &  31.03 &  31.14$\pm$0.11 &   30.97$\pm$0.1  &    31.03$\pm$0.05 &   31.09$\pm$0.04 &  31.06$\pm$0.03\\
NGC  4278 &  31.05 &                 &   30.51$\pm$0.14 &    31.18$\pm$0.16 &   31.09$\pm$0.15 &  31.13$\pm$0.11\\
NGC  1427 &  31.29 &                 &                  &    31.65$\pm$0.27 &   31.64$\pm$0.22 &  31.64$\pm$0.18\\
NGC  4494 &  31.64 &                 &   30.81$\pm$0.08 &    31.13$\pm$0.2  &   31.11$\pm$0.18 &  31.12$\pm$0.14\\
NGC  4365 &   31.7 &  31.94$\pm$0.12 &   31.61$\pm$0.19 &    31.64$\pm$0.23 &   31.72$\pm$0.25 &  31.68$\pm$0.17\\

\tableline
\enddata
\tablenotetext{} {(1) Galaxy : (2) Distance modulus from table 1 : 
(3) SBF distance from Neilsen (1999) : (4) weighted mean from Ferrarese (2000) : (5) V-band GCLF distance (from $\sigma$=1.3 Gaussian GCLF fits) : (6) I-band GCLF distance (from $\sigma$=1.3 Gaussian GCLF fits) : (7) Mean GCLF distance }
\tablenotetext{\dag} {Only galaxies with uncertainties in $<$m-M$>$$_{GCLF}$ $<$ 0.2 mag are listed. The ones
with larger uncertainty are more distant, hence only a small fraction of the GCLF is sampled which can lead to
significant systematic errors. }
\end{deluxetable}

\begin{deluxetable}{llll}
\tighten
\tablenum{7}
\tablecaption{ Elliptical Galaxies: Specific Frequency }
\startdata
\\
\tableline
\tableline
Galaxy    &  N$_{Tot}$ & S$_{N(Loc)}$ &  S$_N$$^\ddag$  \\
(1)	  & (2)	  	& (3)	 & (4) \\	  
\tableline
NGC 4472 &   457$\pm$13   &      1.1$\pm$0.1 &   5.6$\pm$1.7 \\
NGC 4649 &   497$\pm$14   &      1.4$\pm$0.2 &   6.7$\pm$1.4 \\
NGC 4406 &   262$\pm$7    &      0.8$\pm$0.1 &   4.4$\pm$1.2 \\
NGC 4365 &   660$\pm$35   &      2.1$\pm$0.6 &   4.3$\pm$0.6 \\
NGC 5322 &  2352$\pm$1387 &      5.8$\pm$5.6 &              \\
NGC 4494 &   237$\pm$11   &      0.9$\pm$0.3 &   5.2$\pm$1.4 \\
NGC 7626 &  1204$\pm$879  &      2.3$\pm$3.3 &              \\
NGC 5982 &  3699$\pm$4292 &     17.4$\pm$22  &              \\
IC 1459 &   526$\pm$59   &      2.3$\pm$0.8 &              \\
NGC 3610 &   856$\pm$460  &      2.9$\pm$3   &              \\
NGC  584 &   406$\pm$52   &      1.5$\pm$0.6 &              \\
NGC 4621 &   229$\pm$6    &      1.2$\pm$0.3 &   4.8$\pm$1.2 \\
NGC 4552 &   306$\pm$6    &      1.2$\pm$0.2 &              \\
NGC 5813 &   647$\pm$61   &      3.6$\pm$1.8 &    6$\pm$1.5 \\
NGC 4589 &   789$\pm$123  &      5.1$\pm$3.7 &              \\
NGC 4278 &   453$\pm$9    &      3.6$\pm$1   &              \\
NGC 4473 &   217$\pm$7    &      1.5$\pm$0.4 &              \\
NGC 3379 &    74$\pm$2    &      0.5$\pm$0.1 &   1.2$\pm$0.6 \\
NGC  821 &   395$\pm$94   &      2.4$\pm$1.6 &              \\
NGC 3608 &   307$\pm$41   &      2.7$\pm$1.7 &              \\
NGC 4291 &   714$\pm$126  &      6.5$\pm$6.4 &              \\
NGC 1439 &   223$\pm$24   &      2.2$\pm$1.8 &              \\
NGC 1427 &   366$\pm$23   &      3.9$\pm$2   &   4.2$\pm$0.7 \\
NGC 3377 &   126$\pm$5    &      2.2$\pm$0.7 &   2.4$\pm$0.6 \\
NGC 4550 &    92$\pm$5    &      1.4$\pm$0.7 &              \\
NGC 5845 &   134$\pm$21   &      2.6$\pm$1.6 &              \\
NGC 4660 &   176$\pm$6    &      3.6$\pm$1.3 &              \\
NGC 4458 &    83$\pm$11   &      2.5$\pm$3   &              \\
NGC 4486B &   113$\pm$4    &      9.1$\pm$1.5 &              \\
\tableline
Avg$^\dag$	& 	& 2.4$\pm$1.8 (0.4)	 \\
\enddata
\tablenotetext{} {(1) Galaxy :  (2) Projected total number of clusters in the WFPC2 field-of-view : (3) Local specific frequency : (4) Specific Frequency from the compilation of Kissler-Patig (1997) and references therein }
\tablenotetext{\dag} { Avg S$_{N(Loc)}$ of systems with $\delta$S$_{N(Loc)}$ $<$ 3. NGC 4550 is excluded as it is a S0 galaxy.  }
\tablenotetext{\ddag} { The values of S$_N$ have been adjusted to the distance moduli in Table 1.   }
\end{deluxetable}

\begin{deluxetable}{lcc}
\tablenum{8}
\tablecaption{ Elliptical Galaxies: Median Cluster Sizes }
\startdata
    \\
\tableline
\tableline
Galaxy$^\dag$    &  No in  PC & Median Size     \\
	  & 	  &   pc \\
(1)	  & (2)	  	& (3)	  \\	  
\tableline
NGC  3379 &     8 &   1.86  \\
NGC  3377 &    25 &   1.89  \\
NGC  4660 &    22 &   1.55  \\
NGC  4472 &    32 &   2.03  \\
 NGC  4649 &    43 &   2.42  \\
 NGC  4406 &    22 &   2.46  \\
NGC  4621 &    33 &   2.24  \\
NGC  4552 &    39 &   1.88  \\
NGC  4473 &    33 &   2.34  \\
NGC  4550 &    17 &   2.06  \\
NGC  4458 &    11 &   2.29  \\
NGC  4486B &     9 &   2.52  \\
NGC  4278 &    31 &    2.5  \\
NGC  1427 &    38 &   2.44  \\
 IC  1459 &    35 &   1.85  \\
NGC  3608 &    12 &   2.79  \\
NGC   821 &    24 &   3.17  \\
 NGC  4494 &    29 &   3.09  \\
 NGC  4365 &    41 &   2.43  \\
NGC   584 &    24 &   1.87  \\
NGC  1439 &    18 &   2.34  \\
NGC  5813 &    25 &   2.51  \\
NGC  5845 &     9 &   2.05  \\
NGC  4589 &    51 &   2.04  \\
NGC  4291 &    22 &   2.18  \\
NGC  3610 &    38 &    2.3  \\
 NGC  5322 &    34 &   2.57  \\
 NGC  5982 &    11 &    3.4  \\
 NGC  7626 &    25 &   3.32  \\
\tableline
Avg &	&  2.36$\pm$0.4 (0.08) \\
\enddata
\tablenotetext{} {(1) Galaxy : (2) No of candidate objects in the PC chip : 
(3) Median size in parsecs (using distance estimates from Table 1)  }
\tablenotetext{\dag} {Sorted by Table 1 distance. See text for details.}
\end{deluxetable}

\end{document}